%% file: main.tex
\documentclass[sigconf, nonacm]{acmart}

\usepackage{booktabs}
\usepackage{subcaption}
\usepackage{xcolor}
\usepackage{colortbl}
\usepackage{graphicx}
\usepackage{amsmath}
\usepackage{braket}
\usepackage{array}
\usepackage{comment}
\usepackage{tikz}
\usepackage{listings}
\usepackage{seqsplit}
\usepackage{ulem}
\usepackage{algorithm}
\usepackage{algpseudocode}
\usepackage{enumitem}
\usepackage{hyperref}
\usepackage{caption}
\usepackage{tabularx}
\usepackage{multirow}
\usepackage{pifont}
\usepackage{placeins}
\usepackage{tikz}
\usetikzlibrary{arrows.meta,positioning,shapes.geometric,shapes.misc,calc,fit}
\definecolor{darkgreen}{RGB}{0,128,0}
\definecolor{darkblue}{RGB}{0,0,139}
\usepackage{url}
\newcommand{\alginput}[1]{
  \par
  \noindent
  \begin{minipage}{\linewidth}
    \setlength{\parindent}{0pt}
    \raggedright
    \textbf{Input:}~#1
  \end{minipage}
  \par
}
\newcommand{\algoutput}[1]{
  \par
  \noindent
  \begin{minipage}{\linewidth}
    \setlength{\parindent}{0pt}
    \raggedright
    \textbf{Output:}~#1
  \end{minipage}
  \par
}
\newcommand{\cmark}{\ding{51}}
\newcommand{\xmark}{\ding{55}}
\newcolumntype{C}{>{\centering\arraybackslash}X}
\newcommand{\tikzmark}[2]{
    \tikz[overlay,remember picture,baseline]
    \node[anchor=base] (#1) {$#2$};
}
\newcommand{\rev}[1]{{\color{blue}#1}}

\begin{document}
\sloppy
%
\newcommand{\name}{{\sc HamSim}}
\newcommand{\namebf}{\textbf{HamSim}}
\title[Diagonal-Budgeted Trotterization]{Diagonal-Budgeted Trotterization for Efficient Quantum Hamiltonian Simulation}
\input{authors}
%
\input{sections/0-abstract.tex}
%
%

\maketitle
%
\input{sections/1-introduction}
\input{sections/2-background}
\input{sections/3-methodology}
\input{sections/4-experiments}
\input{sections/5-results}
\input{sections/6-related_work}
\input{sections/9.1-future_work}
\input{sections/7-conclusion}
%
\input{sections/8-acknowledgment}
\input{sections/9-appendix}
%
%
\bibliographystyle{ACM-Reference-Format}
\bibliography{references}
\end{document}

%% file: authors.tex
\author{Srikar Chundury}
\email{schundu3@ncsu.edu}
\orcid{0009-0001-8335-9259}
\affiliation{%
  \institution{
    North Carolina State University
  }
  \city{Raleigh}
  \state{NC}
  \country{USA}
}

\author{Blake Burgstahler}
\email{bburgst@ncsu.edu}
\orcid{0009-0005-4249-7857}
\affiliation{%
  \institution{
    North Carolina State University
  }
  \city{Raleigh}
  \state{NC}
  \country{USA}
}

\author{Jiajia Li}
\email{jiajia.li@ncsu.edu}
\orcid{0000-0003-1270-4147}
\affiliation{%
  \institution{
    North Carolina State University
  }
  \city{Raleigh}
  \state{NC}
  \country{USA}
}

\author{In-Saeng Suh}
\email{suhi@ornl.gov}
\orcid{0000-0002-6923-6455}
\affiliation{%
  \institution{
    Oak Ridge National Laboratory
  }
  \city{Oak Ridge}
  \state{TN}
  \country{USA}
}

\author{Frank Mueller}
\email{fmuelle@ncsu.edu}
\orcid{0000-0002-0258-0294}
\affiliation{%
  \institution{
    North Carolina State University
  }
  \city{Raleigh}
  \state{NC}
  \country{USA}
}

%% file: sections/0-abstract.tex
\begin{abstract}
\label{sec:abstract}
Efficient classical simulation of quantum Hamiltonian dynamics is often
bottlenecked by exponential state growth and the overhead of generic
sparse linear algebra. We introduce \emph{diagonal-budgeted
Trotterization}, a structure-aware strategy that decomposes
Hamiltonians into factors preserving diagonal sparsity while tightly
controlling fidelity loss.

Our implementation, \name{}, utilizes a compact diagonal-sparse data
layout and specialized C++/CUDA kernels to bypass the overheads of
generic formats like CSR. By leveraging SIMD vectorization,
multithreading, and GPU acceleration, \name{} achieves high performance
across heterogeneous architectures.
%
Benchmarks on the HamLib suite show that \name{} significantly
outperforms Qiskit-Aer. On CPUs, \name{} attains speedups of
\textbf{$182$--$1{,}269\times$} on optimization instances (TSP, MaxCut)
and \textbf{$4.8$--$841\times$} on physical models (TFIM, Heisenberg).
On GPUs, it achieves up to \textbf{$178\times$} speedup for 12--16
qubit problems.

Unlike traditional Trotterization, \name{} maintains near-perfect
fidelity without requiring exponential steps. This demonstrates that
diagonal-aware numerical kernels provide a scalable
foundation
for high-fidelity classical Hamiltonian simulation.
\end{abstract}

%% file: sections/1-introduction.tex
\section{Introduction}
\label{sec:introduction}
Quantum computing has seen rapid progress
recently,
driven
by foundational algorithms such as Shor's factoring~\cite{Shor},
Grover's search~\cite{Grover1996}, and HHL~\cite{PhysRevLett.103.150502},
as well as variational and hybrid approaches, including the variational
quantum eigensolver (VQE)~\cite{Peruzzo_2014}, the quantum approximate
optimization algorithm (QAOA)~\cite{farhi2014quantumapproximateoptimizationalgorithm,Farhi_2022},
and quantum neural networks (QNNs)~\cite{bergholm2018pennylane}.
These methods have enabled applications in cryptography,
combinatorial optimization, physics and chemistry, machine learning,
and finance~\cite{Shor_1997,farhi2014quantum,Low_2019,Biamonte_2017,Woerner_2019}.
Central to both algorithm development and hardware validation is the
ability to accurately simulate quantum systems.

Hamiltonian simulation plays a particularly important role, as it
models the continuous-time evolution of quantum systems governed by the
Schr{\"o}dinger equation~\cite{Childs_2018}.
Such simulations are critical for studying physical phenomena, including molecular dynamics and many-body systems,
and are especially relevant to today's quantum hardware, which is costly to access,
provides limited coherence times, and is imperfect due to
noise.
Classical Hamiltonian simulation therefore remains a
cornerstone for validating quantum algorithms, benchmarking~\cite{Villalonga_2020},
and exploring algorithmic behavior prior to
hardware execution~\cite{Smelyanskiy:2016ozv}.

A common approach to Hamiltonian simulation is to approximate time
evolution using Trotterization, decomposing the evolution operator into
a sequence of discrete time steps. While higher-order Trotter schemes
can improve accuracy~\cite{SUZUKI1992387}, achieving high fidelity often requires a large
number of time steps, resulting in deep quantum circuits that are
difficult to execute on
hardware. Consequently,
HPC simulation remains indispensable, both for
validation and
exploration of regimes beyond current hardware
~\cite{jones2019quest, li2021svsim, li2020density, full-stack-acceleration_2026}.

However, classical simulation of quantum systems is fundamentally
limited by the exponential growth of the Hilbert space with the number
of qubits. This ``curse of dimensionality'' leads to prohibitive
computational and memory costs, even on modern HPC platforms. Although
parallelism and acceleration help extend feasible problem sizes,
performance is often constrained by the efficiency of sparse linear
algebra kernels, particularly sparse matrix exponentiation and
sparse matrix--vector multiplication~\cite{ge-spmm}.

\begin{figure}[!htbp]
  \vspace*{-0.5\baselineskip}
  \centering
  \includegraphics[width=\linewidth]{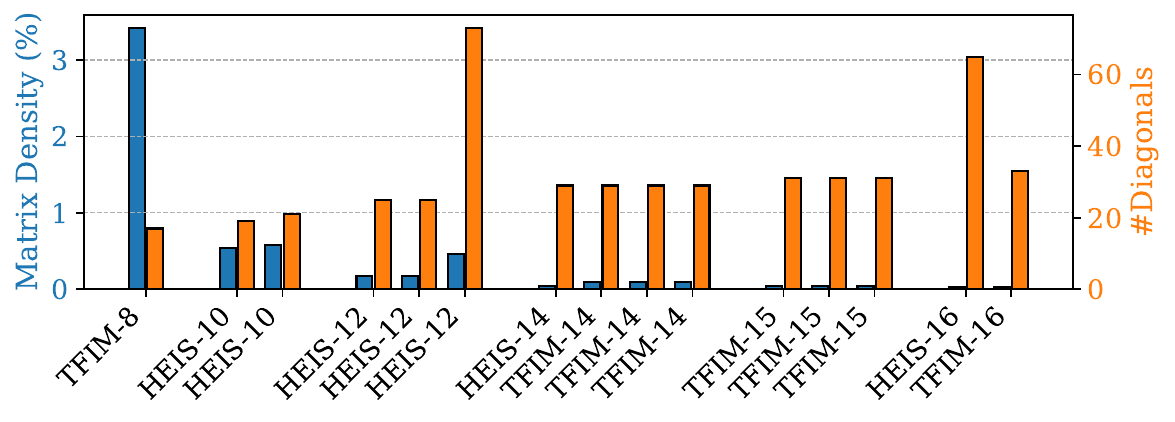}
  \vspace*{-2\baselineskip}
  \caption{Matrix density (left y-axis) and number of active diagonals
    (right y-axis) for selected instances from HamLib~\cite{Sawaya_2024}.
    Identical x-axis labels correspond to distinct Hamiltonians of 
    the same category with the same qubit size.
  }
  \Description{shows matrix density and num diags vs qubit count}
  \label{fig:hamiltonian_diaq_sparsity_analysis}
  \vspace*{-0.5\baselineskip}
\end{figure}

A key observation motivating this work is that problem-Hamiltonians
(that have Hermitian properties when represented using matrices) in
many quantum applications exhibit strong and persistent structural
sparsity. In contrast to generic sparse matrices with irregular
non-zero patterns, such Hermitian operators often concentrate their
non-zero entries along a limited number of diagonals due to their
local interactions and structured operator decompositions (e.g., Pauli
strings). We refer to these as ``active'' diagonals.
Fig.~\ref{fig:hamiltonian_diaq_sparsity_analysis} empirically
illustrates this phenomenon for Hamiltonians drawn from the HamLib
benchmark suite~\cite{Sawaya_2024}: Despite exponential growth in matrix
dimension, both the overall matrix density and the number of active
diagonals grow slowly and remain tightly bounded with qubit count.

\begin{figure}[!htbp]
  \vspace*{-0.5\baselineskip}
  \centering
  \begin{subfigure}[t]{0.495\linewidth}
    \centering
    \includegraphics[width=\linewidth]{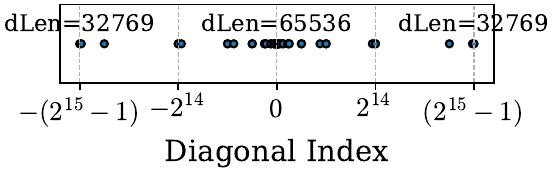}
    \label{fig:heisenberg_16}
  \end{subfigure}
  \hfill
  \begin{subfigure}[t]{0.495\linewidth}
    \centering
    \includegraphics[width=\linewidth]{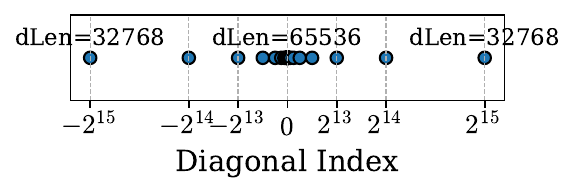}
    \label{fig:tfim_16}
  \end{subfigure}
  \vspace*{-1.5\baselineskip}
  \caption{Distribution of diagonal indices in 16-qubit Hermitian matrices: Heisenberg (left) and TFIM (right).}
  \label{fig:hamiltonian_diaq_sparsity_single}
  \Description{shows the dist. of diagonal indices}
  \vspace*{-0.5\baselineskip}
\end{figure}

Importantly, this diagonal sparsity is not incidental. As shown in
Fig.~\ref{fig:hamiltonian_diaq_sparsity_single}, non-zero diagonals
exhibit clear structure, symmetry, and widening gaps away from the main
diagonal. These properties suggest that much of the computational cost
incurred by existing simulation pipelines stems not from inherent
complexity, but from a mismatch between Hamiltonian structure and the
numerical representations traditionally used in classical simulation.

We address this mismatch by designing a novel time-evolution strategy
and numerical kernels around diagonal sparsity. We introduce
\emph{diagonal-budgeted Trotterization}, a sparsity-aware time-evolution
technique that explicitly controls diagonal growth by selecting a
carefully bounded number of evolution steps whose factors preserve
diagonal structure while tightly controlling fidelity loss.
We realize this technique efficiently in \emph{\name{}}, a diagonal-aware
Hamiltonian simulation approach that uses a compact diagonal-sparse data
layout together with high-performance kernels for sparse matrix
exponentiation, sparse matrix--matrix multiplication, and sparse
matrix--vector multiplication. Implemented in a C++/CUDA framework with
Python bindings, \name{} leverages SIMD vectorization, multithreading,
and GPU acceleration to enable efficient classical Hamiltonian simulation
at scales that are challenging for existing methods.

In summary, this paper makes the following contributions:
\begin{itemize}
  \item We propose \emph{diagonal-budgeted Trotterization}, a
  sparsity-aware time-evolution strategy that bounds the number of
  Trotter steps for high-fidelity Hamiltonian simulation.
  \item We develop a diagonal-aware Hamiltonian simulation approach,
  implemented using a diagonal-sparse data layout and HPC-optimized
  kernels for sparse matrix exponentiation, sparse matrix--matrix
  multiplication, and sparse matrix--vector multiplication.
  \item We implement and evaluate our approach on multi-core CPUs and
  GPUs, demonstrating substantial speedups over
  today's
  simulators
  and strong scaling to larger problem sizes. Across
  HamLib benchmarks, \name{} achieves \textbf{$182$--$1{,}269\times$} speedups on
  TSP and MaxCut and \textbf{$4.8$--$841\times$} on TFIM and Heisenberg on
  CPUs, while \name{}-GPU attains \textbf{$7.8\times$--$37.1\times$} speedups
  on 8--10 qubits and about \textbf{$178\times$} on 12--16 qubit instances,
  all while maintaining near-perfect fidelity.
\end{itemize}

%% file: sections/2-background.tex
\section{Background}
\label{sec:background}
%
%
%
\subsection{Hamiltonian Time Evolution}
\label{subsec:hamiltonian_time_evolution}
Hamiltonian simulation models the continuous-time evolution of quantum
systems governed by the Schr{\"o}dinger equation,
\begin{equation}
	i\hbar\,\frac{d}{dt}\ket{\psi(t)} = H\ket{\psi(t)}
	\label{eq:schrodinger_equation}
\end{equation}
where \(H\) is a Hermitian operator encoding system interactions.
The formal solution is
\begin{equation}
	\ket{\psi(t)} = U(t)\ket{\psi(0)}, \qquad \qquad U(t) = e^{-iHt}.
	\label{eq:time_evolution}
\end{equation}

Hamiltonians in chemistry, condensed matter, and many-body physics are
typically expressed as structured sums of non-commuting terms
\begin{equation}
H = \sum_{j=1}^{m} H_j.
\label{eq:hamiltonian_sum}
\end{equation}

A common classical approximation strategy is Trotterization, which
factorizes the propagator into short-time evolutions,
\begin{equation}
e^{-iHt} \;\approx\;
\Bigg(\prod_{j=1}^{m} e^{-iH_j\,t/r}\Bigg)^{\!r},
\label{eq:trotter_formula}
\end{equation}
where \(r\) is the number of Trotter steps. Increasing \(r\) improves
accuracy but raises computational costs, while higher-order formulas
reduce the step count at the expense of more complex per-step
operators. Reducing the number of steps needed to reach a target
fidelity is therefore critical for both classical simulation performance
and near-term quantum execution.

\subsection{Scalability and Sparsity}
\label{subsec:scalability_and_sparsity}
For an \(n\)-qubit system, both \(H\) and the propagator \(U(t)\) are
\(N \times N\) matrices with \(N = 2^n\). Dense storage requires
\(O(N^2)\) memory, and dense matrix multiplication incurs \(O(N^3)\)
time, making large-scale simulation infeasible since \(N\) grows
exponentially with \(n\). Applying a dense propagator to a
state vector costs \(O(N^2)\) per time step.

In practice, however, physically meaningful Hamiltonians exhibit strong
structural sparsity. As
shown in Fig.~\ref{fig:hamiltonian_diaq_sparsity_analysis}, the number of
active diagonals grows slowly even as
matrix dimension increases
exponentially, reflecting locality, limited interaction range, and the
structured nature of Pauli-operator decompositions. Preserving and
exploiting this diagonal sparsity throughout time evolution is essential
to reduce
memory footprint and computational cost.

Unfortunately, existing simulation pipelines often fail to preserve this
structure
as
they rely on sparsity-oblivious Trotterization and
sparse formats that do not capture diagonal regularity.

\subsection{Limitations of Existing Sparse Formats}
\label{subsec:existing_sparse_formats}
Sparse matrix formats such as Compressed Sparse Row (CSR), Compressed
Sparse Column (CSC), Coordinate (COO), Ellpack (ELL), and diagonal-based
formats like SciPy's DIA representation~\cite{2020_SciPy,saad1990sparskit,li2013smat}
are widely used in numerical computing. These formats are optimized for
arbitrary sparsity patterns and deliver reasonable performance across
general workloads.

Diagonal-based formats organize non-zero entries by offset from the main
diagonal and are efficient when a matrix has \textit{narrow bandwidth},
i.e., when all non-zero entries lie within a small range of diagonal
offsets.
However, when active diagonals lie far from the main diagonal, these
formats must pad each diagonal to the full matrix width, producing large
blocks of unused storage (Tab.~\ref{tab:existing_sparse_formats}).
Additionally, common DIA implementations often fall back to CSR for
matrix operations, reintroducing CSR overheads and limiting achievable
performance.
\input{tables/2.1-sparse_formats_table.tex}
General-purpose formats are therefore ill-suited for Hamiltonian
simulation workloads where sparsity is highly structured, persistent
across time steps, and concentrated along specific diagonal offsets.
This mismatch motivates the development of approaches that
explicitly preserve diagonal structure.

Fig.~\ref{fig:hamiltonian_diaq_sparsity_single} illustrates the
challenge for representative 16-qubit Hamiltonians. Although the number
of active diagonals is small, these diagonals may lie far apart,
creating large empty regions of just zeros. In diagonal-based
formats such as DIA, this leads to substantial padding overhead and
inefficient traversal during matrix operations. As system size
increases, padding costs grow rapidly, and memory traffic begins to
dominate runtime, even though the true number of non-zero entries remains
small.

%% file: tables/2.1-sparse_formats_table.tex
\begin{table}[!htbp]
    \caption{Existing sparse matrix formats vs. \name{}'s layout.}
    \vspace{-0.5\baselineskip}
    \label{tab:existing_sparse_formats}
    \footnotesize 
    \setlength{\tabcolsep}{3pt} 
    \centering
    \begin{tabular}{|c|c|c|}
      \hline
      \multicolumn{3}{|c|}{\textbf{(a) Dense}} \\
      \hline
      \multicolumn{3}{|c|}{
      $
          \begin{pmatrix}
            \tikzmark{diag-top}{a} & 0 & 0 & \tikzmark{diag-b}{b} \\
            0 & c & 0 & 0 \\
            0 & 0 & d & 0 \\
            \tikzmark{diag-e}{e} & 0 & 0 & \tikzmark{diag-bottom}{f} \\
          \end{pmatrix}
      $
      } \\
      & \multicolumn{2}{c|}{
      \begin{tikzpicture}[overlay,remember picture]
        \draw[red,opacity=.2,line width=3mm,line cap=round] (diag-top.center) -- (diag-bottom.center);
        \draw[red,opacity=.2,line width=3mm,line cap=round] (diag-b.center) -- (diag-b.center);
        \draw[red,opacity=.2,line width=3mm,line cap=round] (diag-e.center) -- (diag-e.center);
      \end{tikzpicture}
      } \\
      \hline
      \textbf{(b) CSR} & \textbf{(c) CSC} & \textbf{(d) COO} \\
      \hline
      \begin{tabular}{@{}l@{}}
        Val: $[a,b,c,d,e,f]$ \\
        RPtr: $[0,2,3,4,6]$ \\
        CIdx: $[0,3,1,2,0,3]$
      \end{tabular} &
      \begin{tabular}{@{}l@{}}
        Val: $[a,e,c,d,b,f]$ \\
        CPtr: $[0,2,3,4,6]$ \\
        RIdx: $[0,3,1,2,0,3]$
      \end{tabular} &
      \begin{tabular}{@{}l@{}}
        Val: $[a,b,c,d,e,f]$ \\
        RIdx: $[0,0,1,2,3,3]$ \\
        CIdx: $[0,3,1,2,0,3]$
      \end{tabular} \\
      \hline
      \textbf{(e) ELL} & \textbf{(f) DIA} & \cellcolor[HTML]{FFEEDF}\textbf{(g) \name{}} \\
      \hline
      \begin{tabular}{@{}l@{}}
        Val: $[a,b,c,*,d,*,e,f]$ \\
        CIdx: $[0,3,1,*,2,*,0,3]$
      \end{tabular} &
      \begin{tabular}{@{}c@{}}
        Val:\;
        $\setlength{\arraycolsep}{2pt} 
        \begin{bmatrix}
          a & c & d & f \\
          * & * & * & e \\
          b & * & * & *
        \end{bmatrix}$
      \end{tabular} &
      \cellcolor[HTML]{FFEEDF}
      \begin{tabular}{@{}l@{}}
        Val: $[a,c,d,f,e,b]$ \\
        Off: $[0,-3,3]$
      \end{tabular} \\
      \hline
    \end{tabular}
    \vspace*{-0.5\baselineskip}
\end{table}

%% file: sections/3-methodology.tex
\section{Methodology}
\label{sec:methodology}

This section presents the methodological components underlying
\name{}. We proceed from the algorithmic strategy to the numerical
representation choices and finally to the high-performance execution
kernels that together enable scalable time evolution. We begin with
\emph{diagonal-budgeted Trotterization}, which controls sparsity growth
during time evolution by selecting a timestep that preserves diagonal
structure while meeting a target fidelity. We then describe the
diagonal-aware sparse layout used internally by \name{} to compactly
store and efficiently process the operators arising from this
decomposition. Finally, we present the core diagonal-aware kernels for
sparse matrix exponentiation, matrix--matrix multiplication, and
matrix--vector multiplication that are tailored for
our novel trotterization strategy and data layout forming
the computational backbone of \name{}.

All algorithms are presented in high-level
pseudocode/walk-through
form and are implemented in our C++/CUDA framework, with Python bindings for
integration into existing simulation workflows. 
For clarity, the pseudocode/walk-through focuses on the main algorithmic ideas 
and omits low-level implementation details such as memory
management, norm estimation, diagonal enumeration, and SIMD alignment
logic. These aspects are handled explicitly in the implementation but
are orthogonal to the core algorithmic ideas described here.

\input{sections/3.1-diagonal_budgeted_trotterization.tex}
\input{sections/3.2-the_diaq_format.tex}
\input{sections/3.3-diaq_kernels.tex}
\input{sections/3.4-sparse_hamiltonian_simulation.tex}

%% file: sections/3.1-diagonal_budgeted_trotterization.tex
%
%
\input{images/3.1-diagonal_budget_image.tex}
\subsection{Diagonal-Budgeted Trotterization}
\label{subsec:diagonal_budgeted_trotterization}
Hamiltonian simulation via Trotterization approximates the time-evolution
operator \(e^{-iHt}\) by decomposing the Hamiltonian
\(H=\sum_{j=1}^{m} H_j\) into a product of short-time exponentials.
Classical implementations typically fix the number of Trotter steps
\(r\) \emph{a priori}, based on theoretical error bounds or rough heuristic
choices. These approaches are agnostic to the sparsity structure of the
intermediate operators and often lead to unnecessary fill-in during
exponentiation, undermining the benefits of sparse representations.

We introduce \emph{diagonal-budgeted Trotterization}, a sparsity-aware
time-evolution strategy that explicitly links the Trotter step size to the
diagonal structure of the short-time propagator. As illustrated in
Fig.~\ref{fig:estimate_timesteps_flow_compact}, instead of choosing a
fixed number of steps, the method adaptively selects the smallest \(r\)
such that the per-step operator \(e^{-iH(t/r)}\) satisfies a user-defined
diagonal budget \(D_{\max}\). This constraint limits the growth of
non-zero diagonals and ensures that each short-time propagator remains
efficiently representable in a diagonal-oriented sparse layout.

A key empirical observation is that, for physically meaningful
Hamiltonians, diagonal fill-in increases monotonically as the Trotter
step size grows. Consequently, the feasibility of a given diagonal budget
defines a monotone predicate over \(r\), enabling efficient search
procedures. Given total evolution time \(t\), we determine the minimal
number of steps \(r\) satisfying
\begin{equation}
\mathrm{num\_diags}\!\left(e^{-iH(t/r)}\right) \le D_{\max}.
\label{eq:diagonal_budget_condition}
\end{equation}

This adaptive selection balances sparsity preservation and accuracy:  
Larger steps reduce computation but risk diagonal growth, while smaller
steps maintain sparsity and ensure efficient propagation within \name{}'s
diagonal-centric kernels.

Diagonal-budgeted Trotterization fundamentally differs from traditional
error-driven step selection. Rather than controlling truncation error
directly, it constrains structural complexity, yielding a practical
upper bound on the number of steps required for high-fidelity simulation
while preserving the sparsity patterns common in real Hamiltonians. This
structure-aware decomposition forms the algorithmic foundation upon which
our diagonal-oriented representation and kernels operate.
%
	The overhead of this procedure is incurred as a one-time preprocessing
	step per Hamiltonian and chosen diagonal budget. In practice, we evaluate a
	small number of candidate values of $r$ and
	estimate $\mathrm{num\_diags}(e^{-iH(t/r)})$ to identify the minimal feasible
	step count. This
	search is inexpensive compared to the overall simulation cost and is
	amortized across all subsequent
	time-evolution steps.
%

%% file: images/3.1-diagonal_budget_image.tex
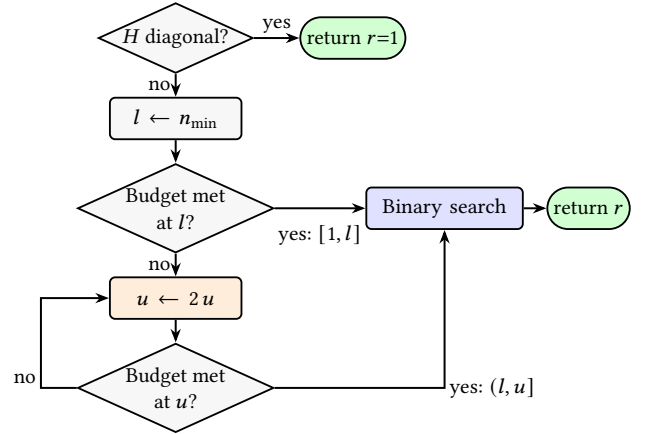
\begin{figure}[!htbp]
\vspace*{-0.5\baselineskip}
\centering
\begin{tikzpicture}[
  font=\small,
  node distance=2.5mm and 4mm,
  arrow/.style={-{Stealth[length=1.8mm]}, line width=0.75pt},
  lab/.style={font=\small},
  every node/.style={inner sep=1.5pt},
  proc/.style={
    rectangle, rounded corners=2pt, draw, line width=0.7pt,
    align=center, inner xsep=2pt, inner ysep=1pt,
    text width=2.4cm, minimum height=5.5mm, fill=gray!6
  },
  dec/.style={
    diamond, draw, line width=0.7pt, aspect=2.2,
    align=center, inner xsep=1pt, inner ysep=0pt,
    fill=gray!6
  },
  term/.style={
    rounded rectangle, rounded rectangle arc length=180,
    draw, line width=0.7pt, align=center,
    inner xsep=2pt, inner ysep=1pt,
    minimum height=5.5mm, fill=green!18
  },
]
\node[dec] (d0) {$H$ diagonal?};

\node[proc, below=3mm of d0, text width=1.6cm] (setL)
  {$l \gets n_{\min}$};

\node[dec, below=3mm of setL] (feasL)
  {Budget met\\at $l$?};

\node[proc, fill=orange!14, below=3mm of feasL, text width=1.6cm] (dbl)
  {$u \gets 2\,u$};

\node[dec, below=3mm of dbl] (feasU)
  {Budget met\\at $u$?};

\node[proc, fill=blue!12, right=12mm of feasL, text width=2cm,
  inner xsep=1pt, inner ysep=0.5pt] (bsearch)
  {Binary search};

\node[term, right=3mm of bsearch] (out)
  {return $r$};

\node[term, right=6mm of d0] (ret1)
  {return $r{=}1$};

\draw[arrow] (d0.south)    -- node[lab, left] {no}  (setL.north);
\draw[arrow] (setL)        -- (feasL);
\draw[arrow] (feasL.south) -- node[lab, left] {no}  (dbl.north);
\draw[arrow] (dbl)         -- (feasU);
\draw[arrow] (bsearch)     -- (out);

\draw[arrow] (d0.east)    -- node[lab, above] {yes} (ret1.west);
\draw[arrow] (feasL.east) -- node[midway, lab, below=1.5mm] {yes: $[1,l]$} (bsearch.west);
\draw[arrow] (feasU) -| node[lab, right] {yes: $(l,u]$} (bsearch.south);

\draw[arrow] (feasU.west) -| node[lab, above left] {no}
  ([xshift=-9mm]dbl.west) -- (dbl.west);

\end{tikzpicture}
\vspace*{-0.5\baselineskip}
\caption{Adaptive timestep selection for diagonal budget $D_{\max}$.
  Given Hamiltonian $H$ and evolution time $t$, find the smallest
  $r$ satisfying Eq.~\eqref{eq:diagonal_budget_condition}.
  $r \!\to\!$ \#steps, $t \!\to\!$ total time,
  $l,u \!\to\!$ bounds on $r$.}
\label{fig:estimate_timesteps_flow_compact}
\Description{Flowchart of Alg.~\ref{alg:estimate_timesteps_by_diagonal_budget} 
with independent early exit, strictly parallel loop logic, and a vertical search routing. 
Vertical spacing is minimized for maximum compactness.}
\vspace*{-1.5\baselineskip}
\end{figure}

%% file: sections/3.2-the_diaq_format.tex
\subsection{Diagonal Representation in \name{}}
\label{sec:diaq_format}

Many problem Hamiltonians and the short-time propagators generated by
diagonal-budgeted Trotterization are sparse with nonzeros concentrated
along a small set of diagonals. To exploit
this structure, \name{} adopts a diagonal-oriented sparse representation
that stores only the active diagonals needed for time evolution, without
the padding overheads incurred by general-purpose diagonal formats such
as SciPy's DIA. Other aspects of the representation are rather similar to DIA
with the key distinction coming from numerical kernels tailored to preserve
diagonal structure during exponentiation and time evolution, as described
in Sec.~\ref{subsec:diaq_kernels}.

The representation follows a diagonal-major layout, where active diagonals are
packed back-to-back into contiguous, SIMD-aligned buffers. Each diagonal
occupies a fixed offset within these buffers, enabling predictable memory
access and efficient vectorized kernels. Tab.~\ref{tab:existing_sparse_formats}
illustrates this layout relative to common sparse formats. Although only
real values are shown for clarity, the representation fully supports
complex-valued matrices.

Formally, a Hamiltonian or propagator in \name{} is stored as
\begin{equation}
  \texttt{DiagRepr}
  = \left\{ \left( d_k,\, v_k^{\mathrm{Re}},\, v_k^{\mathrm{Im}} \right) \mid d_k \in \mathbb{Z} \right\},
  \label{eq:diaq_representation}
\end{equation}
where each diagonal index \(d_k\) corresponds to two contiguous arrays
holding the real and imaginary parts of the diagonal's entries.
Following standard conventions, \(d_k=0\) denotes the main diagonal;
positive (resp.\ negative) values denote superdiagonals (resp.\ subdiagonals).
For an \(N\times N\) operator, the logical length of diagonal \(d_k\) is
\(n_k = N - |d_k|\), and \name{} allocates exactly \(n_k\) elements, with
minimal padding only when required for SIMD alignment.

To maximize performance, real and imaginary components are stored in a
structure-of-arrays (SoA) layout. Although this is not ideal when every
kernel performs coupled complex operations, the vast majority of \name{}'s
inner loops operate on one component at a time (e.g., real accumulations
in intermediate steps), and SoA yields more regular memory access and
significantly better vectorization. This choice also eliminates the need
to store unused imaginary parts for purely real Hamiltonians.

\paragraph{Diagonal-set estimation and preallocation.}
\name{}'s numerical kernels, including matrix exponentiation,
matrix--matrix multiplication, and matrix--vector multiplication, are
implemented out of place. Rather than generating diagonals dynamically,
\name{} performs a lightweight \emph{structural dry run} that propagates
only diagonal offsets. This identifies the exact set of diagonals that
will appear in the output, enabling a single preallocation of contiguous
storage. Fixing the diagonal layout up-front
avoids pointer chasing, dynamic growth, and allocation overhead, all of
which are critical for predictable performance and high SIMD/GPU
efficiency.

\paragraph{Storage implications.}
This diagonal-oriented layout becomes increasingly beneficial at scale.
A \(2^{16}\times 2^{16}\) Hamiltonian with 16 active diagonals at offsets
\(\pm 2^{15}\) contains roughly \(16\times 2^{15}\) actual nonzeros.
\name{}'s layout stores exactly these values, consuming about 8~MB for
double-precision complex data.  
By contrast, SciPy's DIA format pads each diagonal to length \(2^{16}\),
adding another 8~MB of unused entries for the same matrix. Padding costs
grow exponentially with system size and quickly dominate memory usage
for realistic Hamiltonians. While memory savings are not the primary
goal of \name{}, this compact layout offers useful secondary advantages
on memory-constrained devices.

\subsubsection{Contrast with SciPy's DIA Format}
\label{subsec:dia_comparison}

SciPy's DIA format stores all diagonals in a dense
\(n_{\mathrm{diags}} \times N\) matrix, padding every diagonal to the
full dimension \(N\). This design works well for narrow-band matrices,
where most diagonals lie close to the main diagonal. However, many
Hamiltonians arising in physics and quantum algorithms contain active
diagonals far from the main diagonal. As shown in
Fig.~\ref{fig:hamiltonian_diaq_sparsity_single}, representative 16-qubit
Heisenberg and TFIM Hamiltonians contain diagonals at offsets as large
as \(\pm 2^{15}\). Although these diagonals have logical length \(2^{15}\),
DIA pads them to \(2^{16}\), wasting half their entries. At 16 bytes per
double-precision complex entry, this corresponds to 512~KB of padding
\emph{per diagonal}, doubling with each added qubit.
\name{}'s layout avoids this padding entirely by allocating storage proportional
to the true number of nonzeros, independent of diagonal offset. Combined
with its preallocated, SIMD-aligned layout, this provides both memory
efficiency and regular access patterns well suited to the short-time
propagators produced by diagonal-budgeted Trotterization.

Compact storage alone is not sufficient: Hamiltonian simulation also
requires kernels that preserve diagonal structure during exponentiation
and time evolution, which are described next.

%% file: sections/3.3-diaq_kernels.tex
\input{sections/3.3-diagonal_times_diagonal_image.tex}
\subsection{\name{} Kernels}
\label{subsec:diaq_kernels}
Building on the \name{} storage layout, we implement a suite of
diagonal-aware numerical kernels that preserve sparsity and enable
efficient Hamiltonian simulation. Unlike general-purpose sparse
libraries that permit arbitrary fill-in during multiplication, all
kernels in \name{} operate on \emph{fixed diagonal layouts} determined
in advance. This design avoids dynamic sparsity growth, enables
contiguous preallocation of all output buffers, and yields predictable
memory access patterns that map cleanly to SIMD on CPUs and to
regular thread/block decompositions on GPUs.

The kernel suite includes sparse matrix--vector multiplication (SpMV),
sparse matrix--matrix multiplication (SpGEMM), matrix addition and
scaling, Hermitian conjugation, transposition, and sparse matrix
exponentiation (SpME). Together, these primitives provide the core
linear-algebra building blocks required for Trotterized Hamiltonian
evolution and unitary simulation, while preserving diagonal structure
across intermediate computations.
%
%
%
%
%
%
\subsubsection{SpGEMM}
SpGEMM is the most
performance-critical kernel in \name{}, supporting matrix
exponentiation.
Given two diagonally sparse inputs \(A\) and \(B\), the product
\(C = AB\) is formed by enumerating all diagonal pairs
\((d_i,d_j)\) whose offsets satisfy \(r = d_i + d_j\) for a candidate
result diagonal \(r\). This structural phase determines the exact set
of diagonals in \(C\) prior to any numerical work.

Fig.~\ref{fig:spgemm_walkthrough} illustrates the full execution
flow: The structural pass builds \texttt{contributors}[$r$] for every
result diagonal; the numeric pass tiles both the diagonal index space
and the element-range space (\texttt{RES\_BLOCK} and \texttt{ELEM\_BLOCK});
and each tile invokes the
\textcolor{darkblue}{\textproc{MultiplyDiagonals}} microkernel, which streams
contiguous diagonal segments and performs complex fused multiply-add
operations. The figure makes explicit how write-vs-accumulate behavior
is handled and how SIMD parallelism is exposed inside each tile.

Numerical execution is parallelized across result diagonals. When a
diagonal \(r\) has a single contributing pair, its output slice is
written directly; when multiple contributors exist, partial results are
accumulated explicitly. Each diagonal-level multiplication runs in time
linear in the diagonal length.

If both inputs contain only the principal diagonal, SpGEMM reduces to
\(O(N)\). In general, the complexity is
\(O(d_A \cdot d_B \cdot N)\), where \(d_A\) and \(d_B\) denote the
numbers of active (non-zero) diagonals, typically small constants for
Hamiltonians exhibiting structured sparsity.

\subsubsection{SpMV}
SpMV is another dominant kernel during Trotterized time evolution, as each
step applies a diagonally sparse propagator to a state vector. In
\name{}, every diagonal \(d\) contributes an independent shifted
element-wise multiply between the diagonal's contiguous buffer and the
input vector. This yields \(O(d\cdot N)\) work (where \(d\) is the number
of active diagonals, typically a small constant) compared to \(O(N^{2})\)
for dense multiplication. The structural phase (Alg.~\ref{alg:sparse_matrix_vector_product})
partitions each diagonal into element-range segments
\((d,[\textit{start},\textit{end}))\), enabling thread-level
parallelism over tiles, while the numeric phase executes each tile using
a streaming inner loop.

Fig.~\ref{fig:diag_times_vec_schematic} provides a detailed view of
the \textcolor{darkblue}{\textproc{DiagonalTimesVector}} microkernel
that operates inside each segment. The figure highlights how the
diagonal offset induces an affine mapping of the logical index \(i\)
to input/output indices \((j,k)\), allowing uniform vectorized
execution across all diagonal types. Each tile processes elements in
units of \texttt{VEC\_WIDTH} via SIMD loads/stores, applies fused
complex FMA updates, and uses a masked tail to handle partial
vectors. Software prefetching and SoA layout further improve memory
locality, making the kernel bandwidth-efficient and highly
predictable.
The per-tile thread-local results are finally reduced into the global
output vector, completing the
SpMV in linear time.
%
%
\begin{algorithm}[!htbp]
  \caption{\textproc{SpMV}: Sparse Matrix--Vector Product}
  \label{alg:sparse_matrix_vector_product}
  \alginput{Matrix $A$ in \name{}'s layout, vector $x$}
  \algoutput{Result vector $y = A x$}
  \begin{algorithmic}[1]
    \State $N \gets |x|$
    \State $y \gets \textproc{InitializeResultVector}(N, 0)$

    \Statex \textcolor{darkgreen}{\textbf{Phase I (structural):} tile diagonals into element-range tasks}
    \State $\text{tasks} \gets$ partition each $(d, v^{\mathrm{Re}}, v^{\mathrm{Im}}) \in A.\text{diagonals}()$
    \Statex \hspace{12mm} into segments $(d, v^{\mathrm{Re}}, v^{\mathrm{Im}}, \textit{start}, \textit{end})$

    \Statex \textcolor{darkgreen}{\textbf{Phase II (numeric):} parallel over diagonal segments}
    \State allocate thread-local output buffers $\{y_{\text{local}}\}$
    \State \textcolor{red}{\#pragma omp parallel for schedule(static)}
    \For{$(d, v^{\mathrm{Re}}, v^{\mathrm{Im}}, \textit{start}, \textit{end}) \in \text{tasks}$}
      \State \textcolor{darkblue}{\textproc{DiagonalTimesVector}}$(d, v^{\mathrm{Re}}, v^{\mathrm{Im}}, x, y_{\text{local}}, \textit{start}, \textit{end})$
      \Statex \Comment{\textcolor{blue}{Fig.~\ref{fig:diag_times_vec_schematic}}}
    \EndFor

    \Statex \textcolor{darkgreen}{\textbf{Phase III (reduction):} combine partial results}
    \State $y \gets \sum y_{\text{local}}$ \Comment{\textcolor{darkgreen}{thread-local reduction}}
    \State \textbf{return} $y$
  \end{algorithmic}
\end{algorithm}
%
%
\input{sections/3.3-diagonal_times_vector_image.tex}
\subsubsection{SpME}
\begin{figure}[!htbp]
  \centering
  \includegraphics[width=0.80\linewidth]{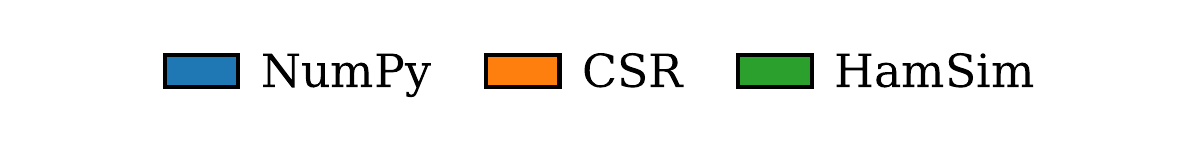}
  \begin{subfigure}[t]{0.235\textwidth}
    \centering
    \includegraphics[width=\linewidth]{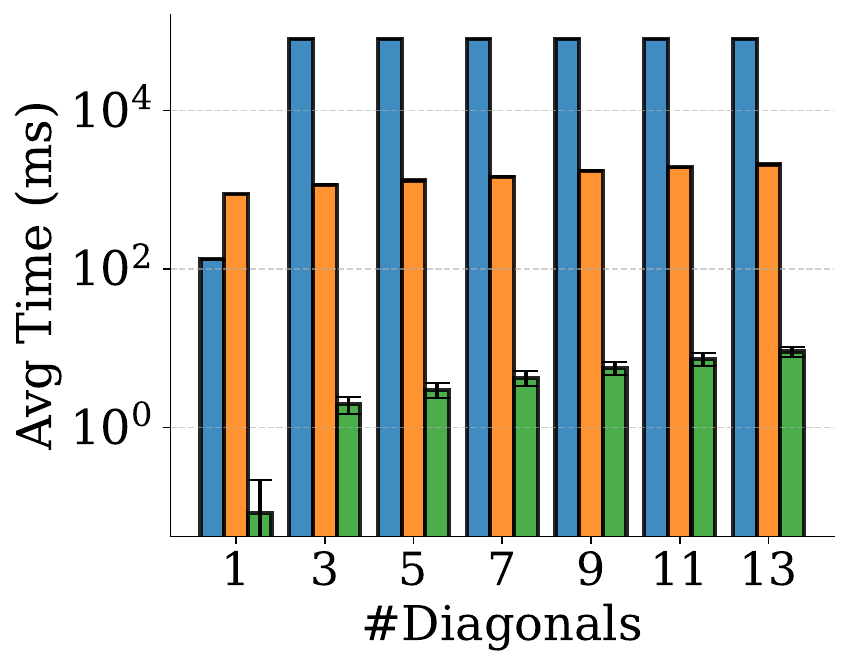}
    \caption{\(t = 0.001\)}
    \label{fig:microbench_t0.001}
  \end{subfigure}
  \hfill
  \begin{subfigure}[t]{0.235\textwidth}
    \centering
    \includegraphics[width=\linewidth]{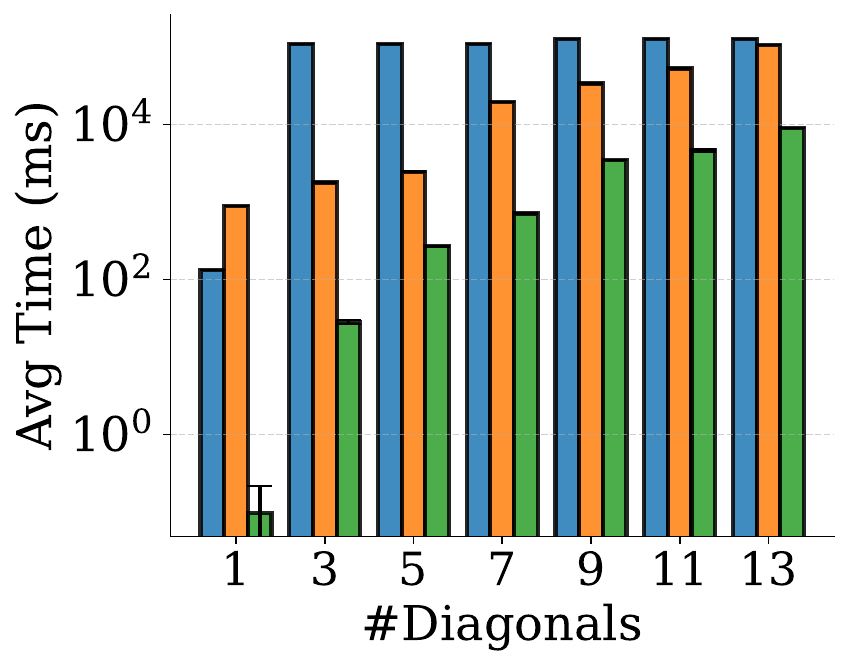}
    \caption{\(t = 1.2\)}
    \label{fig:microbench_t1.2}
  \end{subfigure}
  \hfill
  \begin{subfigure}[t]{0.235\textwidth}
    \centering
    \includegraphics[width=\linewidth]{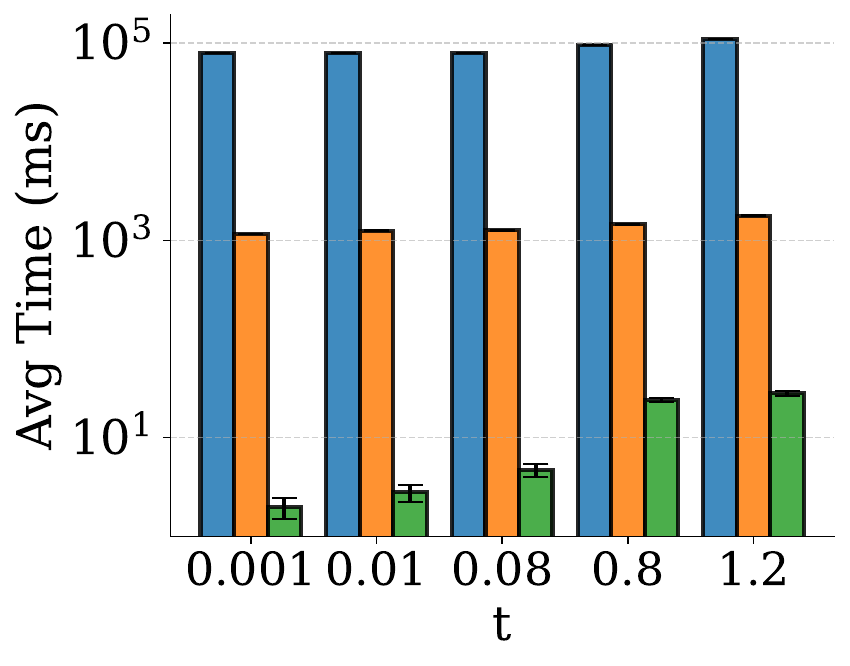}
    \caption{\(d = 3\)}
    \label{fig:microbench_d3}
  \end{subfigure}
  \hfill
  \begin{subfigure}[t]{0.235\textwidth}
    \centering
    \includegraphics[width=\linewidth]{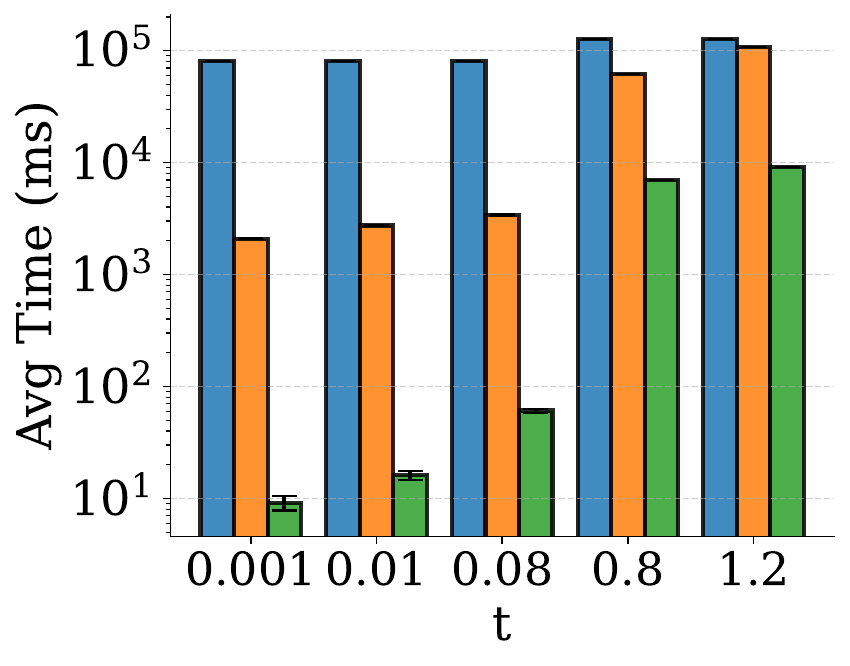}
    \caption{\(d = 13\)}
    \label{fig:microbench_d13}
  \end{subfigure}
  \caption{Performance of \name{}'s matrix exponential on a 12-qubit system for varying $t$ and diagonal counts $d$.}
  \label{fig:exp_comparison}
  \Description{this shows the performance of matrix exp with varying t and d}
\end{figure}

Matrix exponentiation is required to form short-time propagators
of the form \(e^{-iH\Delta t}\). \name{} exploits diagonal
layout to provide fast paths and
scalable approximations. If the input matrix is purely
diagonal, \name{} computes
the exponential exactly via element-wise complex exponentiation
in linear time. For general diagonally sparse Hermitian
matrices, \name{} applies a truncated
Taylor-series expansion (Alg.~\ref{alg:expm_negative_taylor}).
In practice, we use the standard scaling-and-squaring variant
for numerical stability; the squaring phase reduces to
repeated \textproc{SpGEMM} calls and thus benefits
directly from \name{}'s diagonal-aware multiplication.
To prevent fill-in from overwhelming intermediate computations,
we explicitly cap diagonal growth using the diagonal-budgeting strategy
(Sec.~\ref{subsec:diagonal_budgeted_trotterization}).
%
\begin{algorithm}[!htbp]
  \caption{\textproc{SpME}: Matrix Exponential via Taylor Series}
  \label{alg:expm_negative_taylor}
      \alginput{Hermitian matrix $A$ in \name{}'s layout, scalar time $t$, convergence tolerance(\rev{=1e-6}), and maximum iterations}
      \algoutput{Unitary $U = e^{-iAt}$ in \name{}'s layout}
  \begin{algorithmic}[1]
    \State $X \gets A * (-i) * t$
    \State $U, T \gets \text{identityMatrix}(\text{A.rows})$
    \For{$i = 1$ \textbf{to} max\_iterations}
      \State $T \gets \textcolor{darkblue}{\textproc{SpGEMM}}(T, X)$,\quad $T \gets T/i$
      \Comment{\textcolor{blue}{Fig.~\ref{fig:spgemm_walkthrough}}}
      \If{$T.\text{norm}() < \text{tolerance}$} \textbf{break} \EndIf
      \State $U.\text{add}(T)$
    \EndFor
    \State \textbf{return} $U$
  \end{algorithmic}
\end{algorithm}
\paragraph{Performance characteristics.}
Fig.~\ref{fig:exp_comparison} depicts the average execution time
(y-axis in log-scale) of \name{}'s matrix exponential on a 12-qubit
Hamiltonian for varying evolution times and diagonal counts. As
expected, performance is strongest when the number of diagonals is
small and degrades gracefully as intermediate matrices become denser.
For small to moderate time steps and structured Hamiltonians (common in
physical simulation workloads) \name{} maintains both numerical
stability (using a scaled and squared approach to avoid division by
small numbers) and substantial performance advantages over generic
sparse and dense approaches.
%
%
\subsubsection{Implementation details}
All kernels are implemented in C++ with OpenMP multi-threading and
SIMD vectorization (AVX-512, AVX2, or SSE selected at compile
time) for CPUs. For GPUs, CUDA implementations follow the same preallocation strategy
to ensure consistent sparsity layouts across backends.
Memory allocation is alignment-aware. Optional loop blocking
and software prefetching are supported for improved cache
locality.
Python bindings using pybind11~\cite{pybind11} and NumPy~\cite{harris2020array}
converters enable seamless integration with
other Python-based simulation workflows.

Together, these diagonal-aware kernels enable the efficient
realization of
Trotterized Hamiltonian simulation and form the foundation for the
sparse time-evolution methods described next.

%% file: sections/3.3-diagonal_times_diagonal_image.tex
\begin{figure*}[!htbp]
\centering
\scalebox{0.98}{
\begin{tikzpicture}[
  font=\small,
  panel/.style={
    draw, rounded corners=2pt, line width=0.8pt,
    inner xsep=9pt, inner ysep=8pt, fill=gray!6, align=left
  },
  box/.style={
    draw, rounded corners=2pt, line width=0.8pt,
    inner xsep=7pt, inner ysep=6pt, fill=white, align=left
  },
  hot/.style={
    draw, rounded corners=2pt, line width=0.8pt,
    inner xsep=7pt, inner ysep=6pt, fill=orange!10, align=left
  },
  badge/.style={
    draw, rounded corners=1.5pt, line width=0.7pt,
    fill=purple!10, inner xsep=4pt, inner ysep=3pt,
    font=\scriptsize, align=left
  },
  arrow/.style={-{Stealth[length=2.0mm]}, line width=0.8pt},
  sweep/.style={-{Stealth[length=2.0mm]}, line width=0.7pt},
]

\coordinate (O) at (0,0);

\node[panel, anchor=north west, text width=0.97\textwidth, minimum height=2.8cm] (P1) at (O) {};

\node[font=\normalsize\bfseries, anchor=north west]
  at ([xshift=0pt,yshift=0pt]P1.north west)
  {Walkthrough: $C \leftarrow A \times B$ in \name{} \hfill (SpGEMM)};

\coordinate (P1in) at ([xshift=7pt,yshift=-16pt]P1.north west);

\node[box, anchor=north west, text width=155pt] (Abox)
at ([xshift=0pt,yshift=0pt]P1in) {\raggedright
\hspace*{1.5pt}\textbf{Input $A$ (\name{})}\\
\hspace*{1.5pt}Diagonal IDs: $\mathcal{D}_A$\\
\hspace*{1.5pt}SoA buffers: $A^{Re}[\cdot], A^{Im}[\cdot]$\\
\hspace*{1.5pt}Each diagonal is a contiguous slice at fixed \texttt{diagOffset}.
};

\node[box, anchor=north west, text width=155pt] (Bbox)
at ([xshift=10pt,yshift=0pt]Abox.north east) {\raggedright
\hspace*{1.5pt}\textbf{Input $B$ (\name{})}\\
\hspace*{1.5pt}Diagonal IDs: $\mathcal{D}_B$\\
\hspace*{1.5pt}SoA buffers: $B^{Re}[\cdot], B^{Im}[\cdot]$\\
\hspace*{1.5pt}Each diagonal stores only its true logical length.
};

\node[box, anchor=north west, text width=120pt, minimum height=2.1cm] (Cbox)
at ([xshift=10pt,yshift=0pt]Bbox.north east) {\raggedright
\small
\hspace*{1.5pt}\textbf{Output $C$ (preallocated)}\\
\hspace*{1.5pt}Structural pass determines $\mathcal{D}_C$.\\
\hspace*{1.5pt}Allocate $C^{Re}[\cdot],\,C^{Im}[\cdot]$ for all $r\in\mathcal{D}_C$.
};

\node[panel, anchor=north west, text width=0.97\textwidth, minimum height=2.50cm] (P2)
at ([xshift=0pt,yshift=-1.0mm]P1.south west) {};

\node[font=\normalsize\bfseries, anchor=north west]
  at ([xshift=0pt,yshift=0pt]P2.north west)
  {(1) Structural pass: build \texttt{contributors}[$r$]};

\coordinate (P2in) at ([xshift=7pt,yshift=-16pt]P2.north west);

\node[box, anchor=north west, text width=220pt] (Stext)
at ([xshift=0pt,yshift=0pt]P2in) {\raggedright
\textbf{Existence-based enumeration}\\
For each candidate diagonal $r$, scan $d_i\in\mathcal{D}_A$ and set $d_j \leftarrow r-d_i$.\\
If $d_j\in\mathcal{D}_B$, append $(d_i,d_j)$ to \texttt{contributors}[$r$].\\
Finally, $\mathcal{D}_C \leftarrow \text{keys}(\texttt{contributors})$.
};

\node[box, anchor=north west, text width=160pt, minimum height=1cm] (Sbuckets)
at ([xshift=10pt,yshift=0pt]Stext.north east) {\raggedright
\textbf{Buckets (per $r$)}\\[4pt]
\scriptsize
\texttt{contributors}[$r_0$] $=[(d_{a1},d_{b1})]$ \hfill (write)\\
\texttt{contributors}[$r_1$] $=[(d_{a2},d_{b2}),(d_{a3},d_{b3}),\ldots]$ \hfill (accum)
};

\node[panel, anchor=north west, text width=0.97\textwidth, minimum height=2.85cm] (P3)
at ([xshift=0pt,yshift=-1.0mm]P2.south west) {};

\node[font=\normalsize\bfseries, anchor=north west]
  at ([xshift=0pt,yshift=0pt]P3.north west)
  {(2) Numeric pass: 2D tiling over result diagonals and element ranges};

\coordinate (P3in) at ([xshift=7pt,yshift=-16pt]P3.north west);

\node[box, anchor=north west, text width=200pt] (Tdesc)
at ([xshift=0pt,yshift=0pt]P3in) {\raggedright
\textbf{Outer blocking}\\
Tile result diagonals in groups of \texttt{RES\_BLOCK}.\\
For a tile, compute $\max \ell_r$ over its diagonals.\\
Sweep element tiles $[jBase,jEnd)$ in steps of \texttt{ELEM\_BLOCK}.\\
\textbf{Parallelism:} diagonal tiles distributed across threads.
};

\node[box, anchor=north west, text width=170pt, minimum height=2.1cm] (Gbox)
at ([xshift=10pt,yshift=0pt]Tdesc.north east) {};

\node[font=\bfseries, anchor=north west]
  at ([xshift=2pt,yshift=-2pt]Gbox.north west)
  {2D tiles};

\node[anchor=north west, align=left, font=\scriptsize, text width=150pt]
  at ([xshift=8pt,yshift=-12pt]Gbox.north west)
{
\textbf{
\texttt{\quad \quad <RES\_BLOCK} diagonals $\times$ \texttt{ELEM\_BLOCK} elements>\\
Iteration order:}\\[3pt]
1.\ loop diagonal tiles ($r$ groups)\\
2.\ for each tile, sweep element blocks $[jBase, jEnd)$\\
3.\ inside each block, update \texttt{contributors}[r]
};



\coordinate (GridNW) at ([xshift=-70pt,yshift=-21pt]P3.north east);
\coordinate (GridSE) at ($(GridNW)+(56pt,-56pt)$);

\draw[line width=0.7pt, rounded corners=1.5pt] (GridNW) rectangle (GridSE);

\foreach \x in {14,28,42}{
  \draw[line width=0.45pt] ($(GridNW)+(\x pt,0pt)$) -- ($(GridNW)+(\x pt,-56pt)$);
}
\foreach \y in {14,28,42}{
  \draw[line width=0.45pt] ($(GridNW)+(0pt,-\y pt)$) -- ($(GridNW)+(56pt,-\y pt)$);
}

\coordinate (Hnw) at ($(GridNW)+(28pt,0pt)$);
\coordinate (Hse) at ($(GridNW)+(42pt,-14pt)$);
\fill[orange!10, rounded corners=1.0pt] (Hnw) rectangle (Hse);
\draw[dashed, line width=0.7pt, rounded corners=1.0pt] (Hnw) rectangle (Hse);
\node[font=\scriptsize] at ($(Hnw)!0.5!(Hse)$) {tile};

\draw[sweep] ($(GridNW)+(0pt,9pt)$) -- ($(GridNW)+(56pt,9pt)$);
\node[font=\scriptsize, anchor=south]
  at ($($(GridNW)+(0pt,9pt)$)!0.5!($(GridNW)+(56pt,9pt)$)$)
  {diag sweep};

\draw[sweep] ($(GridNW)+(-9pt,0pt)$) -- ($(GridNW)+(-9pt,-56pt)$);
\node[font=\scriptsize, rotate=90, anchor=south]
  at ($($(GridNW)+(-9pt,0pt)$)!0.5!($(GridNW)+(-9pt,-56pt)$)$)
  {elem sweep};

\draw[sweep, line width=0.55pt, opacity=0.45]
  ($(GridNW)+(7pt,-7pt)$) -- ($(GridNW)+(49pt,-7pt)$)
  -- ($(GridNW)+(49pt,-21pt)$) -- ($(GridNW)+(7pt,-21pt)$)
  -- ($(GridNW)+(7pt,-35pt)$) -- ($(GridNW)+(49pt,-35pt)$)
  -- ($(GridNW)+(49pt,-49pt)$) -- ($(GridNW)+(7pt,-49pt)$);

\node[panel, anchor=north west, text width=0.97\textwidth, minimum height=3.8cm] (P4)
at ([xshift=0pt,yshift=-1.0mm]P3.south west) {};

\node[font=\normalsize\bfseries, anchor=north west]
  at ([xshift=0pt,yshift=-1pt]P4.north west)
  {(3) Inside one tile: contributors $\rightarrow$ segment update $\rightarrow$ SIMD};

\coordinate (P4in) at ([xshift=7pt,yshift=-16pt]P4.north west);

\node[box, anchor=north west, text width=190pt] (Rbox)
at ([xshift=0pt,yshift=-3pt]P4in) {\raggedright
\textbf{Pick one result diagonal $r$}\\
Let $\mathcal{L}=\texttt{contributors}[r]$.\\
For each $(d_i,d_j)\in\mathcal{L}$, update $C[r]$ on $[jBase,jEnd)$.\\
\textbf{First:} write \quad \textbf{rest:} accumulate
};

\node[hot, anchor=north west, text width=140pt] (Seg)
at ([xshift=10pt,yshift=0pt]Rbox.north east) {\raggedright
\textbf{Aligned segment update}\\
compute offsets and \textit{jump}\\
clip to $[jBase,jEnd)$,
inner blocking for locality,
dispatch SIMD microkernel
};

\node[box, anchor=north west, text width=100pt] (SIMD)
at ([xshift=10pt,yshift=0pt]Seg.north east) {\raggedright
\textbf{SIMD microkernel}\\
aligned window + vector loop\\
software prefetch
};

\draw[arrow] ([yshift=8mm]Rbox.east) -- ([yshift=8mm]Seg.west);
\draw[arrow] ([yshift=8mm]Seg.east) -- ([yshift=8mm]SIMD.west);

\node[box, anchor=south west, text width=\dimexpr\textwidth-25pt\relax] (Btm)
at ([xshift=7pt,yshift=4pt]P4.south west) {\raggedright
\textbf{Result layout:} $C[r]$ is a contiguous slice in $C^{Re/Im}$ at a fixed \texttt{diagOffset}.\\
\textbf{Update:} first contributor writes; subsequent contributors accumulate.
};

\path[use as bounding box] (P1.north west) rectangle (P4.south east);

\end{tikzpicture}
}
\vspace*{-0.5\baselineskip}
\caption{\textbf{Walkthrough of \name{} SpGEMM ($C{=}AB$).}
\textbf{(1)} A structural pass builds \texttt{contributors}[$r$] via diagonal-index existence checks.
\textbf{(2)} The numeric pass performs 2D tiling over result diagonals (\texttt{RES\_BLOCK}) and element ranges (\texttt{ELEM\_BLOCK}).
\textbf{(3)} Inside a tile, each $r$ iterates contributor pairs and applies an aligned segment update over $[jBase,jEnd)$,
dispatching a SIMD microkernel to write/accumulate into the preallocated $C[r]$ slice.}
\label{fig:spgemm_walkthrough}
\Description{Diagram illustrating the main steps of the SpGEMM algorithm in \name{}.}
\end{figure*}

%% file: sections/3.3-diagonal_times_vector_image.tex
\begin{figure*}[!htbp]
\centering
\scalebox{0.98}{
\begin{tikzpicture}[
  font=\small,
  arrow/.style={-{Stealth[length=2.0mm]}, line width=0.8pt},
  dashedarrow/.style={-{Stealth[length=2.0mm]}, line width=0.8pt, dashed},
  box/.style={
    draw, rounded corners=2pt, line width=0.8pt,
    align=left, inner xsep=4pt, inner ysep=3pt, fill=gray!6
  },
  smallbox/.style={
    draw, rounded corners=2pt, line width=0.8pt,
    align=left, inner xsep=4pt, inner ysep=3pt, fill=white
  },
  hot/.style={
    draw, rounded corners=2pt, line width=0.8pt,
    align=left, inner xsep=4pt, inner ysep=3pt, fill=orange!12
  },
  badge/.style={
    draw, rounded corners=1.5pt, line width=0.6pt,
    fill=purple!12, inner xsep=3pt, inner ysep=2pt,
    font=\small, align=left
  },
  title/.style={font=\small\bfseries},
]

\node[inner sep=0pt] (O) at (0,0) {};

\node[box, anchor=north west, text width=0.29\textwidth] (L)
at ($(O.south west)+(0,-2.5mm)$) {%
\textbf{(1) Diagonal mapping}\par
\smallskip
\textbf{$d<0$ (sub):} $j=i$, $k=i-d$\\
\textbf{$d=0$ (main):} $j=i$, $k=i$\\
\textbf{$d>0$ (super):} $j=i+d$, $k=i$
};

\node[smallbox, anchor=north west, text width=0.29\textwidth] (LP)
at ($(L.south west)+(0,-1mm)$) {%
\textbf{Key property:} affine index functions yield predictable streams and keep
the inner loop uniform across diagonal types.
};

\node[smallbox, anchor=north west, text width=0.29\textwidth] (LHPC)
at ($(LP.south west)+(0,-1mm)$) {%
\textbf{Implementation:}
structure-of-arrays (SoA) layout and fused complex FMA enable
efficient SIMD execution with streaming loads and stores.
};

\node[box, anchor=north west,
      text width=0.655\textwidth, minimum height=4.55cm] (R)
at ($(L.north east)+(4.5mm,0)$) {};

\node[title, anchor=north west] (Rtitle)
at ($(R.north west)+(0pt,0pt)$)
{\textbf{(2) SIMD traversal over a tile $i \in [\textit{start},\textit{end})$}};

\coordinate (Rin) at ($(R.north west)+(10pt,-6pt)$);

\node[smallbox, anchor=north west, text width=0.60\textwidth] (A)
at ($(Rin)+(0,-3.4mm)$) {%
\textbf{Tile kernel:} process $i$ in steps of \texttt{VEC\_WIDTH} using vector loads/stores.\\
Handle the last partial vector with a masked tail when $\textit{end}-i < \texttt{VEC\_WIDTH}$.
};

\node[hot, anchor=north west,
      text width=0.315\textwidth, minimum height=2.3cm] (S)
at ($(A.south west)+(0.118\textwidth+6.0mm,-0.8mm)$) {};

\node[anchor=north west] (Slabel)
at ($(S.north west)+(4pt,-2pt)$) {%
\textbf{SIMD hot loop} $(i{+}{=}\texttt{VEC\_WIDTH})$
};

\coordinate (PhaseShift) at (0,-5.55mm);

\node[smallbox, anchor=north west, text width=0.118\textwidth] (P)
at ($(S.north west)+(-6.0mm-0.118\textwidth,0)+(PhaseShift)$) {%
\textbf{Head}\\
vector\\
$[start,\cdot)$
};

\node[smallbox, anchor=north west, text width=0.118\textwidth] (E)
at ($(S.north east)+(3.0mm,0)+(PhaseShift)$) {%
\textbf{Tail}\\
masked\\
($<\texttt{VEC\_WIDTH}$)
};

\draw[arrow] (P.east) -- (S.west);
\draw[arrow] (S.east) -- (E.west);

\node[badge, anchor=south west, text width=0.145\textwidth] (Sblk)
at ($(S.south west)+(1.6mm,1.6mm)$) {%
\textbf{Tiling (2D):}
tasks: $(d,[start,end))$\\
\texttt{TILE\_LEN = BLOCK\_SIZE}
};

\node[badge, anchor=south east, text width=0.145\textwidth] (Spf)
at ($(S.south east)+(-1.6mm,1.6mm)$) {%
\textbf{Prefetch:}\\
\texttt{x/y index + PREFETCH\_DIST}\\
$v^{Re/Im},x^{Re/Im},y^{Re/Im}$
};


\node[smallbox, anchor=north west, text width=0.595\textwidth] (K)
at ($(Rin)+(0,-3.7cm)$) {%
\textbf{Update:} $y[k] \mathrel{+}= v[i]\cdot x[j]$ (complex FMA)\quad
\textbf{SoA:} $(v^{Re},v^{Im}), (x^{Re},x^{Im}), (y^{Re},y^{Im})$.
};

\draw[dashedarrow] (L.east) to[out=8, in=180] (A.west);
\draw[dashedarrow] (LHPC.east) to[out=0, in=180] (K.west);

\end{tikzpicture}
}
\vspace*{-0.5\baselineskip}
\caption{\textbf{Schematic of the \textproc{DiagonalTimesVector} kernel.}
A diagonal offset induces an affine mapping $(j,k)$ from logical position $i$ to
input/output indices, enabling a uniform streaming update. In the implementation,
each diagonal is partitioned into $(d,[start,end))$ tiles (2D tasking), and each tile
executes a SIMD hot loop with vector loads/stores and a masked tail to handle
partial vectors; software prefetching overlaps memory latency with fused complex FMA updates.
\textbf{GPU:} we support a single-diagonal element-parallel kernel and a multi-diagonal row-parallel kernel.}
\label{fig:diag_times_vec_schematic}
\Description{Diagram illustrating the main steps of the DiagonalTimesVector kernel in \name{}.}
\end{figure*}

%% file: sections/3.4-sparse_hamiltonian_simulation.tex
\subsection{Efficient Sparse Hamiltonian Simulation}
\label{subsec:sparse_hamiltonian_simulation}
The diagonal-aware kernels described in
Sec.~\ref{subsec:diaq_kernels}, together with the diagonal-budgeted
trotterization strategy introduced in
Sec.~\ref{subsec:diagonal_budgeted_trotterization}, enable an end-to-end
Hamiltonian simulation pipeline that preserves sparsity throughout time
evolution.

Given a Hermitian (problem-Hamiltonian) $H$ and total evolution time $T$, the goal
is to approximate $e^{-iHT}x$ for an initial state vector $x$ without
ever materializing dense matrices or allowing uncontrolled fill-in.
\name{}'s layout stores the problem-Hamiltonian only with a
small number of diagonals, reflecting the
physical interactions.

When $H$ is purely diagonal, the time evolution operator can be applied
exactly in $O(N)$ time via element-wise complex exponentiation.
For more general diagonally sparse Hamiltonians, we rely on a
Trotterized time evolution scheme in which the total time $T$ is split
into $n$ steps of size $\Delta t = T/n$. The key challenge is to choose
$n$ large enough to control fill-in during exponentiation, but not so
large as to introduce unnecessary computational overhead.

Rather than fixing $n$ a priori, \name{} determines it adaptively using
the diagonal-budgeted search procedure introduced in
Sec.~\ref{subsec:diagonal_budgeted_trotterization}.
This routine selects
the smallest $n$ such that the short-time propagator
$e^{-iH\Delta t}$ remains within a user-specified diagonal budget
$D_{\max}$.

Alg.~\ref{alg:hamiltonian_simulation} summarizes the resulting
simulation workflow. Once $n$ is determined, a sparse unitary
$U \approx e^{-iH\Delta t}$ is constructed using the \textproc{SpME} kernel.
The state is then evolved by
repeated application of $U$ using the \textproc{SpMV} kernel.
Because both kernels operate
directly on the \name{} layout and respect the diagonal budget, all
intermediate operators remain sparse and efficiently computable.

%
%
\begin{algorithm}[!htbp]
  \caption{Diagonal-Budgeted Hamiltonian Simulation}
  \label{alg:hamiltonian_simulation}
  \alginput{Hermitian matrix $A$ in \name{} layout, initial state $x$, total evolution time $T$, diagonal budget $D_{\max}$}
  \algoutput{Final state $x_T \approx e^{-i A T} x$}
  \begin{algorithmic}[1]
    \State $n \gets \textcolor{darkblue}{\textproc{EstimateTimesteps}}(A, T, D_{\max})$
      \Comment{\textcolor{blue}{Fig.~\ref{fig:estimate_timesteps_flow_compact}}}
    \State $\Delta t \gets T / n$
    \State $U \gets \textcolor{darkblue}{\textproc{SpME}}(A, \Delta t)$
      \Comment{Scaled sparse exponential, \textcolor{blue}{Alg.~\ref{alg:expm_negative_taylor}}}
    \For{$i = 1$ \textbf{to} $n$}
      \State $x \gets \textcolor{darkblue}{\textproc{SpMV}}(U, x)$
      \Comment{\textcolor{blue}{Alg.~\ref{alg:sparse_matrix_vector_product}}}
    \EndFor
    \State \textbf{return} $x$
  \end{algorithmic}
\end{algorithm}

This diagonal-budgeted approach achieves two objectives simultaneously,
which are difficult to satisfy jointly using conventional
error-driven Trotterization.
First, it bounds memory usage and arithmetic cost by explicitly limiting
the number of active diagonals in each operator. Second, it avoids
overly conservative timestep choices dictated by worst-case dense
approximations. As shown in Sec.~\ref{sec:results}, this combination
allows \name{} to achieve effectively unit fidelity relative to dense
baselines while delivering substantial performance gains for structured
Hamiltonians common in quantum simulation workloads.
%
%

%% file: sections/4-experiments.tex
\section{Evaluations}
\label{sec:experiments}
%
%
\subsection{Setup}
\label{sec:experimental_setup}
Experiments were conducted on a high-performance CPU-based
platform, an AMD EPYC 8124P 16-Core Processor with
192\,GB of DDR5 4800\,MHz ECC memory. Programs are compiled using GCC~12.3
with the \texttt{-O3} optimization flag. To leverage parallelism,
OpenMP is enabled across all runs, using all 32 hardware
threads with simultaneous multithreading (SMT) on this platform.

GPU experiments were performed on a single NVIDIA H100 NVL PCIe Gen5 GPU
(94\,GB HBM3, SM~9.0).
All GPU kernels are implemented in CUDA and compiled with \texttt{nvcc}
using \texttt{-O3} and the default device-side optimizations. The kernels
operate directly on the diagonal-sparse \name{} layout: for sparse
matrix--vector multiplication we use a specialized single-diagonal kernel
when the matrix contains only one diagonal, and a row-parallel kernel for
the general multi-diagonal case; for sparse matrix--matrix multiplication,
each result diagonal is assigned to a thread block, and threads within the
block iterate over diagonal positions and accumulate contributions from all
contributing diagonal pairs.

Unless otherwise stated, kernels are launched with 256 threads per block
and a one-dimensional grid sized to cover all active rows (SpMV) or result
diagonals (SpGEMM). Complex-valued data are stored in a
structure-of-arrays layout (separate real and imaginary arrays) to
improve global-memory coalescing. Reported GPU times correspond to kernel
execution time after explicit synchronization.

\begin{table}[!htbp]
  \centering
  \small
  \caption{Comparison of time-evolution methods.}
  \vspace*{-0.5\baselineskip}
  \label{tab:evolution_methods}
  \begin{tabularx}{\linewidth}{
      l
      >{\hsize=0.5\hsize\arraybackslash}X
      >{\hsize=0.25\hsize\arraybackslash}X
      >{\hsize=0.25\hsize\centering\arraybackslash}X
    }
      \toprule
      \textbf{Method} & \textbf{Compute} & \textbf{Memory} & \textbf{Adaptive} \\
      \midrule
      \textbf{NumPy}              & $O(N^3)$                      & $O(N^2)$             & \xmark \\
      \textbf{CSR}              & $O(T \cdot \mathrm{nnz}(H))$  & $O(\mathrm{nnz}(H))$ & \xmark \\
      \textbf{\texttt{expm\_multiply}}    & $O(T \cdot \mathrm{nnz}(H))$  & $O(\mathrm{nnz}(H))$ & \xmark \\
      \textbf{Qiskit--Aer}      & $O(T \cdot L \cdot N)$        & $O(N)$               & \cmark \\
      \textbf{\name}        & $O(T \cdot D \cdot N)$        & $O(D\,N)$            & \cmark \\
      \bottomrule
  \end{tabularx}
  \begin{minipage}{\linewidth}
    \raggedright
      \footnotesize
      $n$: number of qubits; $N{=}2^n$: Hilbert-space dimension;
            $H$: Hermitian Hamiltonian; $\mathrm{nnz}(H)$: nonzeros in $H$; 
            $L$: number of terms in $H$; $D$: active diagonals; 
            $t$: final evolution time; $T$: timesteps; $\Delta t = t/T$: step size.
            “Adaptive” indicates adaptive timestep selection.
  \end{minipage}
  \vspace*{-0.5\baselineskip}
\end{table}
\subsection{Baselines}
\label{subsec:baselines}
We compare \name{} against state-of-the-art approaches
under identical inputs and precision. A high-level compute/memory
summary is presented in Tab.~\ref{tab:evolution_methods}.
  All baselines use OpenMP/SIMD (CPU) or
  CUDA (GPU) parallelism with a consistent
  number of threads to ensure fair comparison.
%
\input{tables/benchmarks_small.tex}

\noindent\textbf{NumPy:}
Constructs the full $N\times N$ propagator or state explicitly in
double precision. It is useful as a fidelity reference but quickly becomes
infeasible as memory grows quadratically with Hilbert-space size.

\noindent\textbf{SciPy--CSR:}
Operates on matrices in Compressed Sparse Row format without diagonal
awareness. It supports general sparse kernels but does not exploit
Hamiltonian structure.

\noindent\textbf{SciPy--DIA:}
Uses SciPy's built-in DIAgonal sparse matrix format and kernels. It is optimized for
diagonal-sparse matrices but has a few differences from \name{}'s layout as discussed in
Sec.~\ref{subsec:dia_comparison}.
  To add, while SciPy's DIA format is designed for diagonal-sparse
  matrices, it does not maintain a fully diagonal-aware execution
  path. For operations such as matrix exponentiation, it typically
  performs format conversions (e.g., to CSR), after which computation
  proceeds using general-purpose sparse kernels.
  SciPy does not exploit diagonal-wise blocking,
  SIMD/vectorized fused kernels, or
  structure-aware accumulation across diagonals.

\noindent\textbf{\texttt{SciPy--expm\_multiply}:}
Applies $e^{-iHt}$ directly to a vector without forming the full
propagator. It is efficient for moderately sparse matrices, but not
tuned for diagonal sparsity.

\noindent\textbf{Qiskit--Aer (1st-order Trotter):}
Simulates time evolution by sequentially applying Hamiltonian terms to
a full state-vector, whose dimension doubles with each qubit. The number
of steps is chosen to match \name{}'s slicing, i.e., runtime scales with
both the number of terms and the statevector length, but this often
incurs noticeable fidelity loss at the same step budget.
Our GPU results compare against state-of-the-art GPU
simulation provided by \textbf{Qiskit-Aer-GPU} using the
cuQuantum~\cite{bayraktar2023cuquantum} (specifically, cuStateVec) library.

\noindent\namebf{}:
Our novel diagonal-budgeted sparse Hamiltonian simulation technique on CPU (\name{}) and GPU (\name{}-GPU).

\subsection{Metrics}
\label{subsec:metrics}
We evaluate our approach using the following metrics:

\noindent\textbf{Simulation Time:} Total wall-clock time is measured, averaged over $10$ runs, with
whiskers of bar charts providing the standard deviation. All 
methods
are executed under identical input and
software configurations.

\noindent\textbf{State Fidelity:} To assess numerical accuracy, 
we compute
the fidelity between the quantum state produced by \name{} and a
reference solution obtained using a dense simulation (using NumPy, when
feasible).

\noindent\textbf{Scalability and Memory Efficiency:} We assess scalability
by increasing the number of qubits in each benchmark.
Existing methods can achieve high fidelity only for small problem sizes,
but become infeasible as the system size grows due to memory or runtime constraints.
In contrast, methods that scale to larger qubit counts typically suffer significant
fidelity loss.
\name{} uniquely combines high fidelity with scalability,
enabling accurate simulation of large quantum systems beyond the
reach of other approaches.

\subsection{Benchmarks}
\label{subsec:benchmarks}
We evaluate \name{} on a diverse set of Hamiltonian simulation
problems drawn from the
HamLib~\cite{Sawaya_2024} benchmark suite. We select four
representative problems spanning multiple quantum domains: the
\textbf{Transverse Field Ising Model (TFIM)} and the
\textbf{Heisenberg model} from condensed matter physics, and the
\textbf{Traveling Salesman Problem (TSP)} and \textbf{MaxCut} from
discrete and binary optimization.
HamLib provides each Hamiltonian as Pauli strings with complex
coefficients, which we load into whatever representation each simulator
expects: dense matrices for NumPy, CSR/DIA for SciPy,
\texttt{SparsePauliOp} for
Qiskit-Aer, and directly into \name{}'s diagonal-sparse target format
without any conversion overhead.

Tab.~\ref{tab:benchmarks} summarizes our benchmarks' categories, applications,
and structural properties (sparsity and diagonal count).

%% file: tables/benchmarks_small.tex
\begin{table}[!htbp]
  \centering
  \small
  \caption{HamLib benchmarks used in this study; Labels match the figures.}
  \vspace*{-0.5\baselineskip}
  \label{tab:benchmarks}
  \begin{tabularx}{\linewidth}{
      l
      l
      >{\centering\arraybackslash}p{0.9cm}
      >{\centering\arraybackslash}p{1.4cm}
      >{\centering\arraybackslash}p{2.0cm}
    }
      \toprule
      \textbf{Application} & \textbf{Label} & \textbf{Qubits} & \textbf{\#Diagonals} & \textbf{Sparsity (\%)} \\
      \midrule
      \multirow{6}{*}{TSP}
        & TSP-1 & 8  & 1 & 99.6094 \\
        & TSP-2 & 8  & 1 & 99.6094 \\
        & TSP-3 & 15 & 1 & 99.9969 \\
        & TSP-4 & 15 & 1 & 99.9969 \\
        & TSP-5 & 16 & 1 & 99.9985 \\
        & TSP-6 & 16 & 1 & 99.9985 \\
      \midrule
      \multirow{7}{*}{Heisenberg}
        & HEIS-1 & 6  & 13 & 93.7500 \\
        & HEIS-2 & 6  & 13 & 93.7500 \\
        & HEIS-3 & 8  & 25 & 97.3145 \\
        & HEIS-4 & 8  & 25 & 97.3267 \\
        & HEIS-5 & 10 & 19 & 99.4629 \\
        & HEIS-6 & 10 & 19 & 99.4629 \\
        & HEIS-7 & 14 & 27 & 99.9548 \\
      \midrule
      \multirow{8}{*}{MaxCut}
        & MAXCUT-1 & 10 & 1 & 99.9390 \\
        & MAXCUT-2 & 10 & 1 & 99.9386 \\
        & MAXCUT-3 & 12 & 1 & 99.9840 \\
        & MAXCUT-4 & 12 & 1 & 99.9834 \\
        & MAXCUT-5 & 14 & 1 & 99.9957 \\
        & MAXCUT-6 & 14 & 1 & 99.9958 \\
        & MAXCUT-7 & 16 & 1 & 99.9989 \\
        & MAXCUT-8 & 16 & 1 & 99.9989 \\
      \midrule
      \multirow{6}{*}{TFIM}
        & TFIM-1 & 7  & 15 & 93.7500 \\
        & TFIM-2 & 8  & 17 & 96.5820 \\
        & TFIM-3 & 9  & 19 & 98.0469 \\
        & TFIM-4 & 10 & 21 & 98.9258 \\
        & TFIM-5 & 14 & 29 & 99.9084 \\
        & TFIM-6 & 16 & 33  & 99.9741 \\
      \bottomrule
  \end{tabularx}
  \vspace*{-0.5\baselineskip}
\end{table}

%% file: sections/5-results.tex
\begin{figure*}[!htbp]
  \vspace{0.5\baselineskip}
  \centering
  \makebox[\textwidth][c]{%
    \includegraphics[width=0.75\textwidth]{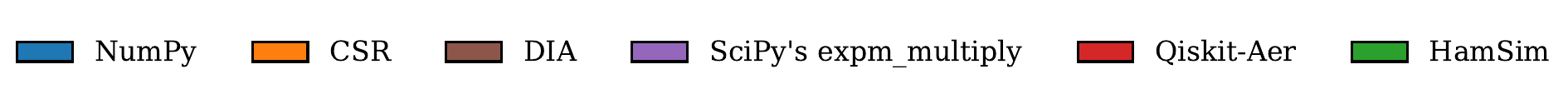}%
  }
  \begin{subfigure}[t]{0.49\textwidth}
    \centering
    \includegraphics[width=\linewidth]{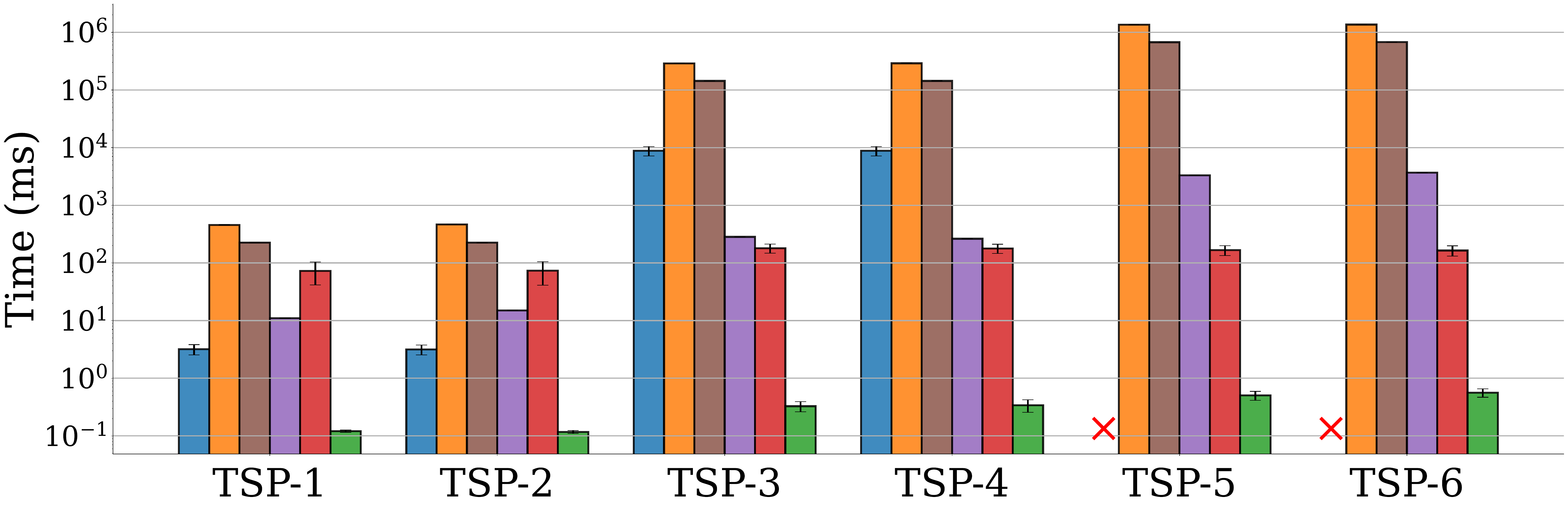}
    \caption{}
    \label{fig:tsp_siena}
  \end{subfigure}\hfill
  \begin{subfigure}[t]{0.49\textwidth}
    \centering
    \includegraphics[width=\linewidth]{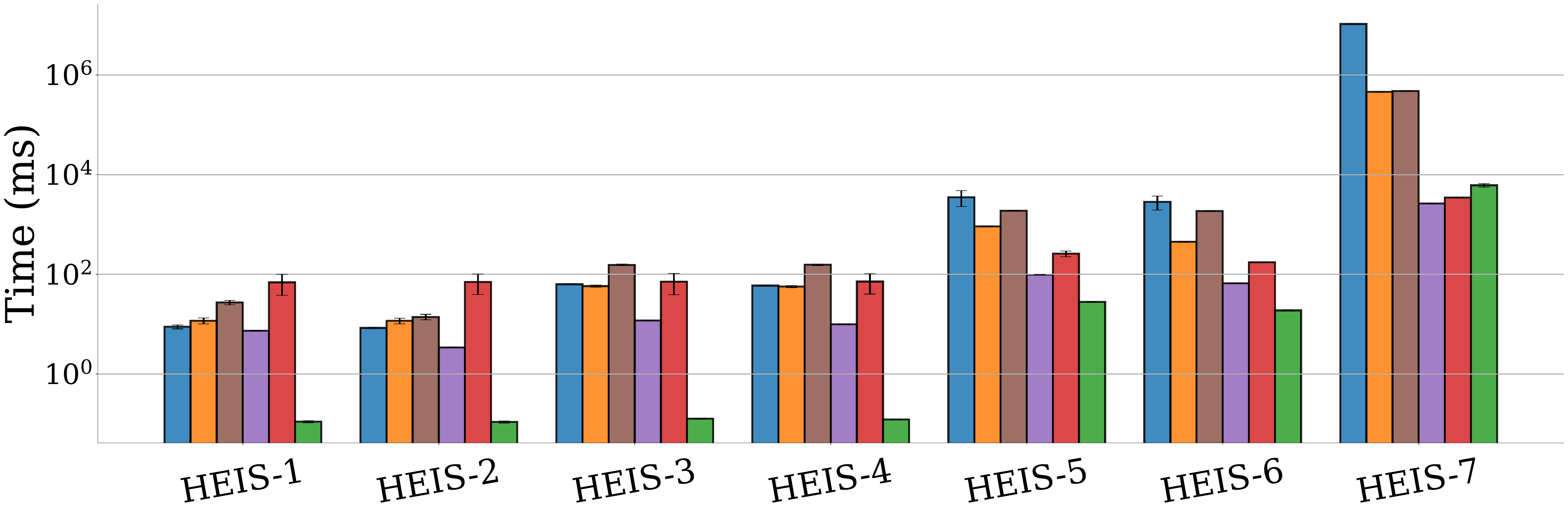}
    \caption{}
    \label{fig:heisenberg_siena}
  \end{subfigure}
  \begin{subfigure}[t]{0.49\textwidth}
    \centering
    \includegraphics[width=\linewidth]{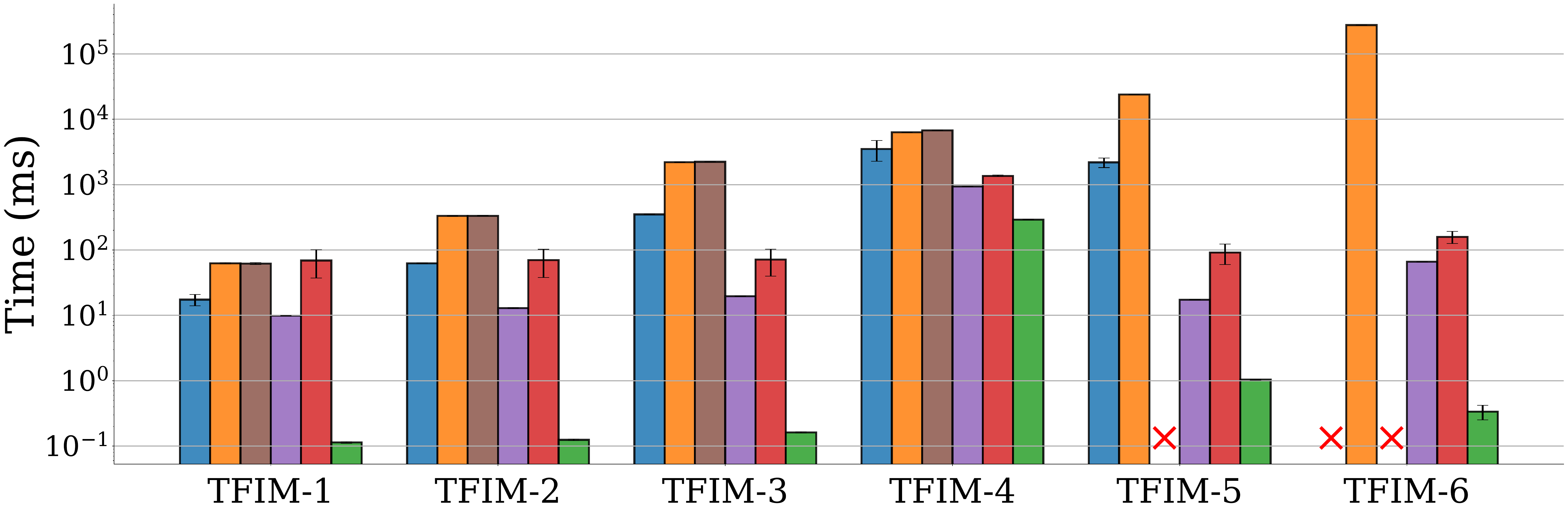}
    \caption{}
    \label{fig:tfim_siena}
  \end{subfigure}\hfill
  \begin{subfigure}[t]{0.49\textwidth}
    \centering
    \includegraphics[width=\linewidth]{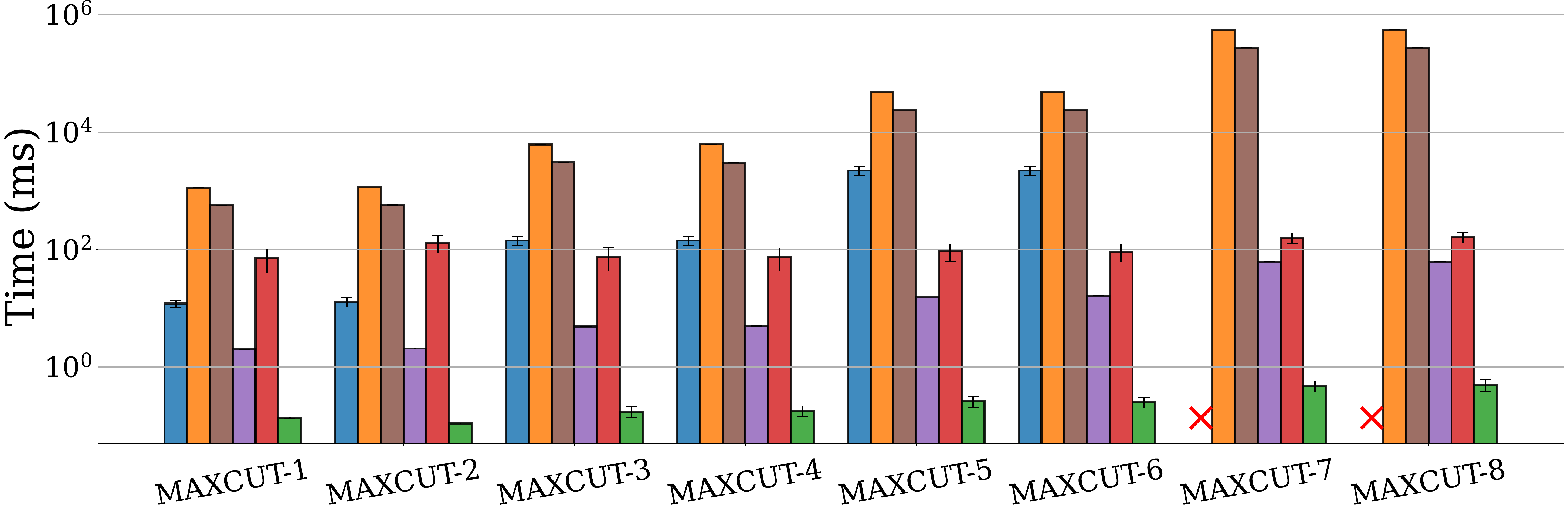}
    \caption{}
    \label{fig:maxcut_siena}
  \end{subfigure}
  \caption{
    Benchmark runtimes across methods for (a) TSP,
    (b) Heisenberg, (c) TFIM, and (d) MaxCut.\\
    \textbf{Note:} For TFIM and Heisenberg, the Qiskit-Aer simulator
    results are shown without fidelity constraints for a fair runtime
    comparison.  Their fidelities are significantly lower, as the
    number of steps is chosen to match \name{}'s settings. A red X indicates
    that the method did not produce a result due
    to memory-allocation failure (out of memory).
  }
  \label{fig:benchmark_results}
  \Description{all benchmark results}
  \vspace{-0.5\baselineskip}
\end{figure*}
\section{Results and Discussion}
\label{sec:results}
\subsection{Performance Comparison against Baselines}
\label{sec:results_and_analysis}
Fig.~\ref{fig:benchmark_results} reports simulation times (log-scale
$y$) over HamLib benchmarks~\cite{Sawaya_2024} along $x$. We compare
\name{} (green; our work) against Qiskit-Aer (red), NumPy (blue; dense,
$100\%$-fidelity baseline), CSR (orange), DIA (brown), and
\texttt{expm\_multiply} (purple). Below, $r$ denotes the number of
timesteps used to apply the propagator (i.e., the number of time-slicing
steps).

\noindent\textbf{TSP (Fig.~\ref{fig:tsp_siena}):}
On the $8$-qubit \texttt{enc-stdbinary} benchmark, \name{}
achieves \textbf{637--652$\times$} speedups over Aer and
\textbf{90--125$\times$} over \texttt{expm\_multiply}, while matching the
dense reference exactly (fidelity $\approx 1.000000$). At $15$ qubits,
\name{} further widens the gap to \textbf{315--323$\times$}
(vs.\ Aer) and \textbf{442--487$\times$} (vs.\ \texttt{expm\_multiply}),
with fidelity again at dense level. On the $16$-qubit \texttt{enc-unary}
instances, \name{} remains sub-millisecond as baselines
diverge, delivering \textbf{(3.6--3.9)$\times10^{3}$} over
\texttt{expm\_multiply}, \textbf{182--193$\times$} over Aer, and
\textbf{(1.4--1.5)$\times10^{6}$} over CSR; in this regime, baselines do
not provide fidelity (or stable accuracy), whereas \name{} matches dense
references whenever fidelity is reported.

\noindent\textbf{MaxCut (Fig.~\ref{fig:maxcut_siena}):}
Across $10$--$14$ qubits, \name{} converts second-scale sparse
simulation into sub-millisecond execution, yielding
\textbf{543--1{,}269$\times$} speedups over Aer and
\textbf{(7.8$\times10^{3}$)--(1.2$\times10^{5}$)} over CSR, while also
outperforming \texttt{expm\_multiply} by \textbf{16--43$\times$}. When
fidelity is available, \name{} reproduces the dense solver to numerical
precision. At $16$ qubits, \name{} sustains strong advantage
with \textbf{70--74$\times$} over \texttt{expm\_multiply} and
\textbf{187--201$\times$} over Aer; baselines provide no dense-level
fidelity guarantee at this scale.

\noindent\textbf{Heisenberg and TFIM (Figs.~\ref{fig:heisenberg_siena},~\ref{fig:tfim_siena}):}
For $6$--$8$ qubits, \name{} delivers dramatic improvements:
\textbf{628--739$\times$} (Heisenberg) and \textbf{506--841$\times$}
(TFIM) over Aer, and \textbf{34--98$\times$} (Heisenberg) and
\textbf{91--124$\times$} (TFIM) over \texttt{expm\_multiply}. Crucially,
\name{} maintains dense-level fidelity ($\approx 1.000000$), while Aer’s
accuracy collapses at the same $r$ (Heisenberg: as low as $0.18$; TFIM:
$0.03$--$0.37$). As propagators densify and $r$ increases, speedups
tighten but \name{} preserves accuracy: for the $10$-qubit TFIM case
($r{=}1120$), \name{} achieves \textbf{3.2--4.8$\times$} over
\texttt{expm\_multiply} and \textbf{1.4--1.6$\times$} over Aer while
maintaining fidelity $\approx 1.000000$. For $10$-qubit Heisenberg
($r{=}71$--$107$), \name{} continues to provide \textbf{32--68$\times$}
speedups over dense baselines with strict fidelity preservation. At the
largest sizes ($14$--$16$ qubits), \name{} completes all simulations
reliably, while multiple baselines either underperform severely or fail
outright and provide no fidelity assurances.
  For HEIS-7, Aer may appear competitive in runtime at the chosen $r$;
  however, this occurs at substantially lower fidelity, whereas \name{}
  maintains near-exact accuracy. Achieving comparable
  fidelity with Aer would require increasing $r$, leading to significantly
  higher runtime.
%
\begin{figure}[!htbp]
    \begingroup
    \centering
    \makebox[0.5\textwidth][c]{%
      \includegraphics[width=1.0\linewidth]{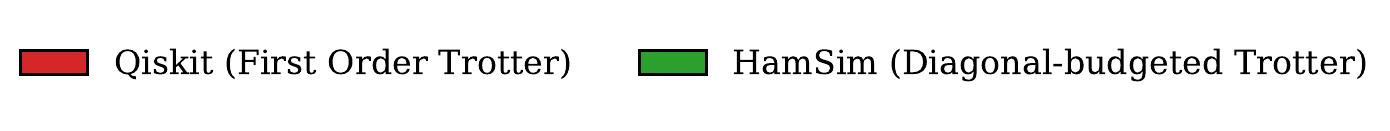}%
    }%
    \vspace*{-0.75\baselineskip}
    \begin{subfigure}[t]{0.49\linewidth}
      \centering
      \includegraphics[width=\linewidth]{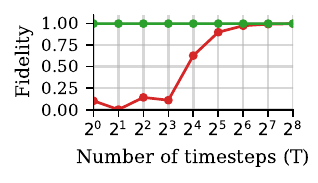}
  \vspace*{-1.5\baselineskip}
      \caption{}
      \label{fig:fid_heis_1}
    \end{subfigure}\hfill
    \begin{subfigure}[t]{0.49\linewidth}
      \centering
      \includegraphics[width=\linewidth]{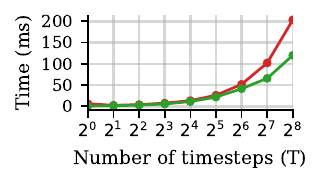}
  \vspace*{-1.5\baselineskip}
      \caption{}
      \label{fig:fid_heis_2}
    \end{subfigure}
    \caption{Heisenberg (12-qubit): (a) Fidelity vs. number of timesteps; (b) Simulation time vs. number of timesteps.}
    \label{fig:fidelity_study}
    \endgroup
    \Description{sensitivity study between fidelity and simulation time for Heisenberg 12 qubit}
\end{figure}
\subsection{Fidelity}
As seen in Fig.~\ref{fig:fidelity_study}, techniques other than Qiskit-Aer
reach $\ge\!99.99\%$ fidelity if/when they complete. \name{} completes with far 
fewer steps (often a single
one) because
\textproc{EstimateTimesteps} (Fig.~\ref{fig:estimate_timesteps_flow_compact}) 
selects the smallest $\Delta t$ that preserves diagonal
sparsity of $U(\Delta t)$ while meeting a fidelity target.
This diagonal-aware budgeting is key: It
keeps the sparse Taylor propagator cheap (SpGEMM over a
banded pattern). In contrast, there exist Hamiltonians where
$U(\Delta t)$ densifies rapidly. Exact diagonal methods can slow in
this regime. Reducing $\Delta t$ (more, smaller steps) restores
diagonal sparsity and performance without sacrificing accuracy. Aer
also reduces $\Delta t$ but does not maintain diagonal structure per
slice, so both per-step cost (due to fill-in) and step count grow.
Dense and CSR baselines require full fill-in throughout.
\begin{figure}[!htbp]
    \vspace*{-0.5\baselineskip}
    \begingroup
    \centering
    \makebox[0.5\textwidth][c]{%
      \includegraphics[width=1.0\linewidth]{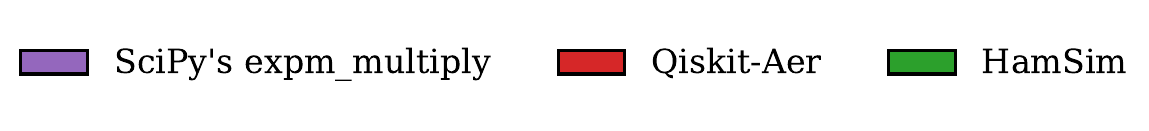}
    }
    \begin{subfigure}[t]{0.49\linewidth}
      \centering
      \includegraphics[width=\linewidth]{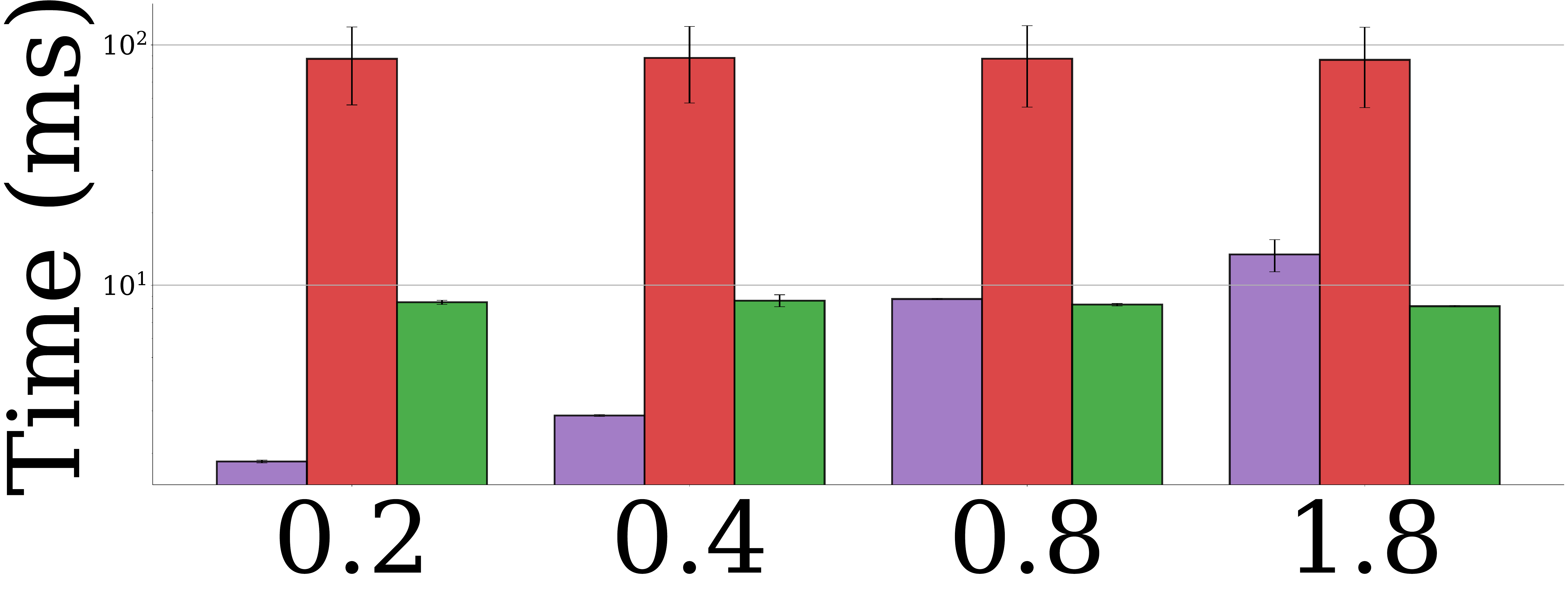}
      \caption{}
      \label{fig:tsp_8}
    \end{subfigure}
    \begin{subfigure}[t]{0.49\linewidth}
      \centering
      \includegraphics[width=\linewidth]{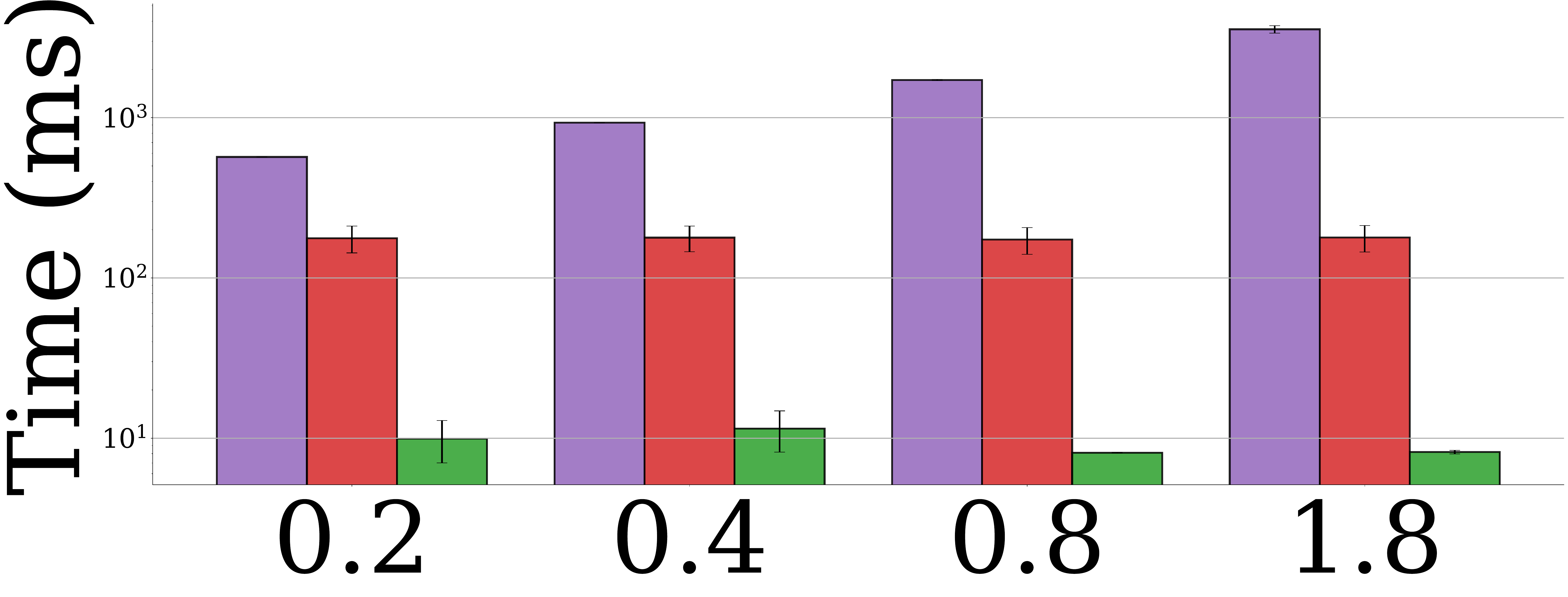}
      \caption{}
      \label{fig:tsp_16}
    \end{subfigure}
    \caption{Simulation time vs. final time $T$ for TSP for (a) 8-qubits and (b) 16-qubits.}
    \label{fig:tsp_sensitivity_combined}
    \endgroup
    \Description{TSP}
    \vspace*{-0.5\baselineskip}
\end{figure}
\subsection{Sensitivity to Final Time}
\name{} maintains high fidelity and speed across
$T\!\in\!\{0.2,0.4,0.8,1.8\}$
(Fig.~\ref{fig:tsp_sensitivity_combined}) but requires significantly
fewer steps than Aer. NumPy and CSR remain accurate but slow, and
degrade more steeply as $T$ increases. For instance, on a fixed
$12$-qubit Hamiltonian (Fig.~\ref{fig:exp_comparison}), performance is flat and stable for
small $t$ and degrades gracefully as $t$ grows and the number of
populated diagonals increases (as seen from $3$ to $13$ diagonals).
The benefit of diagonal sparsity is largest when few diagonals are
active. As the band widens, the gap to dense-style kernels narrows,
which is precisely the behavior observed in TFIM/Heisenberg at larger
sizes.
\begin{figure}[!htbp]
    \begingroup
    \centering
    \makebox[0.5\textwidth][c]{%
      \includegraphics[width=1.0\linewidth]{images/thread_legend.pdf}%
    }
    \includegraphics[width=\linewidth]{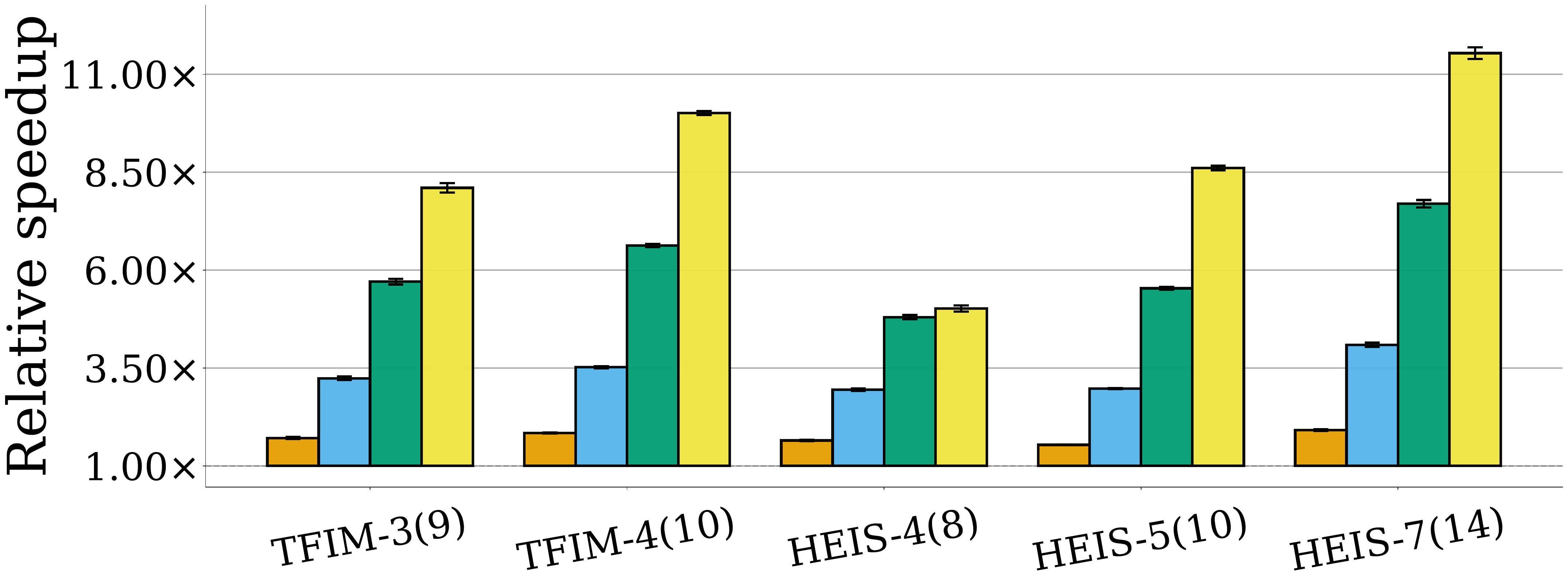}
    \caption{OpenMP strong-scaling of \name{} on our Platform (16-core),
      baseline is single-thread.}
    \label{fig:omp_scaling}
    \endgroup
    \Description{OpenMP strong scaling}
    \vspace*{\baselineskip}
\end{figure}
\subsection{Multithreading and SIMD}
Fig.~\ref{fig:omp_scaling} shows that \name{} strong-scales up to 16
threads on our platform, delivering $8$-$11\times$ over the
single-threaded baseline on multi-diagonal Hamiltonians. Scaling
tapers beyond 16 because the node switches to SMT (32 logical threads
on 16 cores) as sibling threads share core execution units and caches,
so additional threads add little throughput or sometimes even slightly regress.

Our OpenMP design parallelizes across nonzero
diagonals. Single-diagonal inputs (e.g, TSP) 
expose little outer-loop
parallelism due to the simplicity of a single 
diagonal.

We also prototyped
threading within a diagonal using per-thread accumulators and a final
reduction. In our bandwidth-sensitive SpMV/SpGEMM kernels this added
synchronization and memory traffic and was slower
end-to-end, so \name{} adopts diagonal-level threading with SIMD
inside the inner loop, orthogonal to OpenMP, and accelerates each
diagonal's inner loop, adding a further $1.4$--$2\times$ on larger
instances ($\gtrsim 10$ qubits), with smaller gains on small ones
where the loop trip count approaches the vector width.

\begin{figure*}[!htbp]
    \begingroup
    \centering
    \makebox[\textwidth][c]{%
      \includegraphics[width=0.4\textwidth]{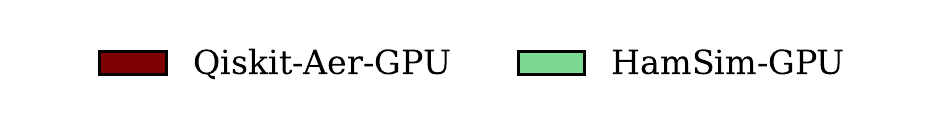}%
    }%
    \vspace*{-\baselineskip}
    \includegraphics[width=\linewidth]{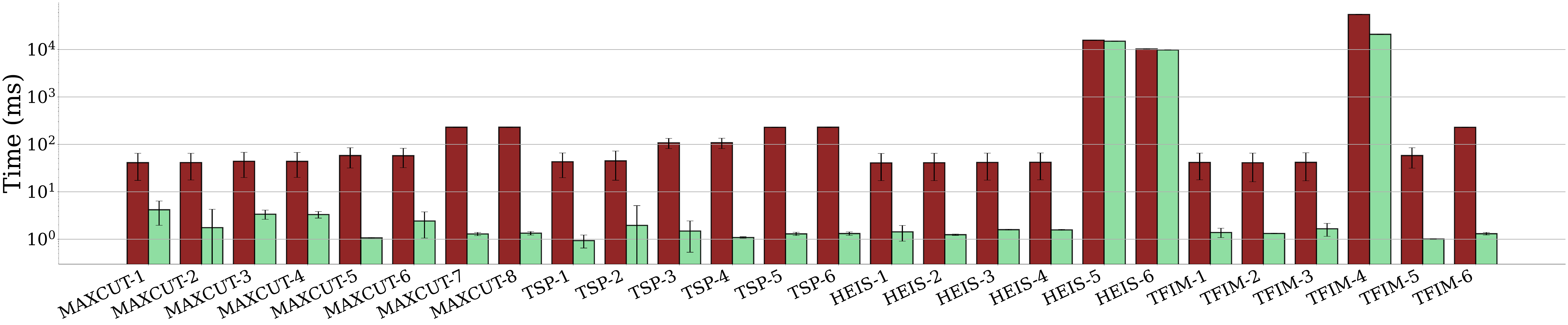}
    \vspace*{-\baselineskip}
    \caption{GPU performance of \name{}-GPU against Qiskit-Aer-GPU (cuQuantum underneath)}
    \label{fig:gpu_results}
    \endgroup
    \Description{gpu performance of \name{}-GPU against qiskit-aer-gpu}
\end{figure*}
\subsection{GPU Performance}
Fig.~\ref{fig:gpu_results} compares \name{}'s GPU kernels against
Qiskit-Aer-GPU (which uses NVIDIA's cuQuantum library under the hood).
For smaller problems (8--10 qubits, across TSP, MaxCut, TFIM, Heisenberg),
\name{}'s GPU kernels deliver substantial speedups over Aer-GPU, typically in the
range of \textbf{$7.8\times$--$37.1\times$}.
For larger problems (12--16 qubits), \name{}-GPU
continues to outperform Aer-GPU, achieving \textbf{$45\times$--$178\times$} speedups
across benchmark families, with peak gains of \textbf{$178.6\times$} on several
16-qubit TSP and MaxCut instances.

In terms of accuracy, like our CPU results, \name{}-GPU also
consistently attains near-perfect fidelity across all benchmarks,
including the largest 14-16 qubit cases.  Aer-GPU, in contrast,
displays substantial fidelity variation as the system size and Trotter
structure grow. For example, on TFIM-1 (7 qubits), Aer-GPU reports a
fidelity of just \textbf{0.0066} despite taking comparable simulation
time, whereas \name{}-GPU remains effectively at $\approx 1.0$.  Even
in the one instance where Aer-GPU's runtime approaches ours, namely
HEIS-5 (10 qubits, 14.99\,s vs.\ 15.75\,s), \name{}-GPU maintains its
high accuracy (fidelity 1.0 vs.\ 0.8243).

On GPU, \name{} operates directly on the diagonal-sparse \name{} layout and avoids
materializing dense intermediates. For SpMV we use a specialized single-diagonal
kernel when $D{=}1$ and a row-parallel kernel for the general multi-diagonal case.
For SpGEMM, each result diagonal is assigned to a thread block whose threads
iterate over diagonal positions and accumulate contributions from all contributing
diagonal pairs.

%% file: sections/6-related_work.tex
\section{Related Work}
\label{sec:related_work}
Several quantum simulators, such as
SparQSim~\cite{sparsqs_2025} and GraFeyn~\cite{grafeyn_2024},
accelerate simulation by tracking only non-zero components of
quantum state vectors. While effective for shallow or
structured circuits, their performance degrades as circuits
deepen or state vectors densify --- limitations that are
particularly evident in Hamiltonian simulation. In contrast,
\name{} focuses on the Hamiltonian itself, decoupling performance
from state-vector sparsity and enabling efficient time evolution.

Other approaches, including Azure Sparse
Simulator~\cite{azure_sparse_2025} and
related work~\cite{leveraging_state_sparsity_2021}, compress
state vectors with dictionary-based or CSR-like
representations.
Diagonal formats have likewise been applied to gate-based
state-vector simulators~\cite{diaq_2024}, yielding performance
gains on large circuits.
There exist sparse libraries like cuSPARSE~\cite{cusparse},
MKL Sparse BLAS~\cite{intel_onemkl},
and other block formats~\cite{1454783}
but none preserve the long-range
diagonal offsets, leading to padding or
scattered accesses. Hence,
these simulators/methods often incur overhead from format
conversions and are not tuned for
Hamiltonian simulations. In contrast,
\name{} maintains diagonal locality and directly
supports matrix exponentiation and matrix-vector
multiplication natively for Hamiltonian simulation.

Theoretical advances in sparse Hamiltonian simulation, such as
Taylor-series-based methods~\cite{childs2012hamiltonian,Berry_2015,berry_improvement_2013} and
quantum walk techniques~\cite{berry_sparse_2005,
childs_star_2010}, provide asymptotic complexity bounds
	on the cost of simulating $e^{-iHt}$ in terms of
	$t$, $\epsilon$, and $\|H\|$, with complexities polylogarithmic in
	$1/\epsilon$ and near-linear in $t\|H\|$, but are formulated in
	abstract algorithmic models and do not directly apply to
	high-performance classical simulation.
	In particular, they assume oracle access to sparse Hamiltonians and do not account for concrete
	data layouts, memory movement, or practical kernel-level implementations.
\name{} draws on these insights but addresses practical
challenges through
diagonal-budgeted Trotterization.

Adaptive-step product formulas have been explored as
an alternative to fixed-step Trotter schemes. Zhao et al.\ propose
ADA-Trotter~\cite{zhao2023makingtrotterizationadaptiveenergyselfcorrecting},
which uses a feedback loop
to select the largest Trotter step size consistent with bounded
deviations in conserved quantities, effectively performing a
search over feasible step sizes. Ostmeyer~\cite{Ostmeyer_2023}
surveys optimized Suzuki--Trotter decompositions~\cite{SUZUKI1992387,Ostmeyer_2023}
and highlights
adaptive stepping and embedded error estimators as promising
directions. Ikeda and Fujii~\cite{Ikeda_2024}
develop a precision-guaranteed method that estimates Trotter error
via embedded decompositions and adjusts the step size accordingly.
\name{} differs from these techniques by targeting
\emph{structure-preserving} time evolution under a diagonal-sparsity
budget, using step selection to maintain diagonal locality and
fidelity rather than only controlling global approximation error.

General-purpose libraries like
Quantuloop~\cite{quantuloop_2024}, QuTiP~\cite{qutip_2012}, and
QSW\_MPI~\cite{qsw_mpi_2017} support sparse simulation using
CSR, CSC, or bitwise formats. However, they do not exploit
structured diagonal sparsity, and often rely on dense matrix
exponentiation or
fixed-format sparse kernels. Thus, comparisons against dense
(NumPy, Qiskit-Aer) and various sparse baselines are more representative for
evaluating \name{}.

%% file: sections/9.1-future_work.tex
\section{Future Work}
\label{sec:future_work}

Several directions build naturally on this work.
The most immediate kernel-level improvement is fusing the Taylor expansion of
$e^{-iH\Delta t}$ directly into the state-vector update, eliminating the
intermediate matrix and cutting memory traffic at large qubit counts.
Fourth-order Suzuki-Trotter is also worth exploring: it cuts per-step error,
which could permit a coarser diagonal budget for the same accuracy and thus
fewer total steps, though it adds extra DiaQ SpME sub-steps per iteration
and how that trade-off plays out under a fixed budget remains an open question.
On the usability side, $D_{\max}$ is currently set by the user via
\texttt{PERCENT\_DIAGONALS}; a short profiling pass over a few values of $r$
would automate this choice.
For scalability, MPI distribution is the natural path beyond single-node
memory limits: diagonal $d$ maps element $i$ to $i{+}d$, so inter-node
communication reduces to a fixed-offset boundary exchange.
Time-dependent Hamiltonians $H(t) = H_0 + f(t)H_1$ fit naturally since the
budget search on $H_0$ and $H_1$ amortizes across all time steps.
Open-system Lindblad dynamics extend cleanly as well, since local jump
operators are diagonal-sparse Kronecker products.
Finally, DiaQ has no quantum-specific structure: lattice force matrices,
circulant signal-processing operators, regular-degree graph Laplacians, and
site-local DMRG/MPS operators all share the same diagonal-sparse pattern,
making \textit{libdiaq} a candidate backend well beyond Hamiltonian simulation.

%% file: sections/7-conclusion.tex
\section{Conclusion}
\label{sec:conclusion}
%
This work demonstrates that the efficiency of classical Hamiltonian
simulation depends on the alignment between numerical kernels and
physical operator structure. While many quantum Hamiltonians exhibit
persistent diagonal sparsity, conventional simulators use
sparsity-oblivious representations that fail to exploit this property.

We introduced \emph{diagonal-budgeted Trotterization}, a strategy
that explicitly bounds diagonal growth to maintain high simulation
fidelity. Our implementation, \name{}, realizes this via a
\emph{domain-specific sparse layout} and HPC-optimized kernels
providing efficient execution on multicore CPUs and GPUs.

Evaluating HamLib benchmarks, \name{} achieves CPU speedups of
\textbf{$182$--$1{,}269\times$} for optimization problems and
\textbf{$4.8$--$841\times$} for physical models. On GPUs, \name{}
attains up to \textbf{$178\times$} speedup. Crucially, \name{}
maintains \textbf{near-perfect fidelity} ($\approx 1.0$) in regimes
where existing simulators suffer from accuracy loss or prohibitive
latency.

This work underscores the value of \emph{structure-driven
design} in scientific computing. \name{} provides a foundation for
accelerating near-diagonal operators, offering a scalable path for
exploiting problem-specific structure in large-scale numerical
workloads.
%

%% file: sections/8-acknowledgment.tex
\begin{acks}
\label{sec:acks}

This work was supported in part by NSF CISE-2217020, CISE-2316201,
OMA-2120757, PHY-1818914, PHY-2325080 and DOE DE-SC0025384.
This research used resources of the Oak Ridge Leadership
Computing Facility at the Oak Ridge National Laboratory, which is
supported by the Office of Science of the U.S. Department of Energy
under Contract No. DE-AC05-00OR22725.
{\it Notice}: This manuscript has been authored by UT-Battelle, LLC, under contract DE-AC05-00OR22725 with the US Department of Energy (DOE). The US government retains and the publisher, by accepting the article for publication, acknowledges that the US government retains a nonexclusive, paid-up, irrevocable, worldwide license to publish or reproduce the published form of this manuscript, or allow others to do so, for US government purposes. DOE will provide public access to these results of federally sponsored research in accordance with the DOE Public Access Plan (https://www.energy.gov/doe-public-access-plan).

\end{acks}

%% file: sections/9-appendix.tex

\section*{Appendix}
\subsection{Benchmark Details}
\input{tables/benchmarks}

Tab.~\ref{tab:benchmarks_large} extends Tab.~\ref{tab:benchmarks} with the full \textbf{HDF5File/Key} paths used as input to \texttt{benchmark\_apps/app2\_hamlib\_latest.py}.

\subsection{Artifact: libdiaq}

The \texttt{libdiaq} C++ library is available at \url{https://github.com/srikarchundury/diaq}.
Prerequisites: cmake $\geq$ 3.20, a C++17-capable compiler (gcc $\geq$ 12 or
Apple Clang $\geq$ 14), and optionally CUDA $\geq$ 11 for GPU kernels.

\textbf{Build and test:}

\begin{lstlisting}[frame=single, basicstyle=\small\ttfamily, breaklines=true]
# Clone with submodules (pybind11, parallel-hashmap)
$ git clone --recurse-submodules \
    https://github.com/srikarchundury/diaq.git
$ cd diaq

# Build CPU library and test binary
$ ./build.sh cpu -DMAKE_TESTS=ON
\end{lstlisting}

\textbf{Python quick start (SpMV):}

\begin{lstlisting}[frame=single, basicstyle=\small\ttfamily, breaklines=true, language=Python]
import numpy as np
import diaq as dq

H_np = np.array([[1,0,0,0],[0,-1,2,0],
                 [0,2,-1,0],[0,0,0,1]],
                 dtype=np.complex128)
psi = np.array([1, 0, 0, 0], dtype=np.complex128)

H      = dq.from_numpy(H_np)
psi_dq = dq.from_numpy_vector(psi)
y_dq   = dq.spMV(H, psi_dq)
y      = dq.to_numpy_vector(y_dq)  # y == H @ psi
\end{lstlisting}

\subsection{Artifact: HamSim}

The HamSim benchmarking suite is available at \url{https://github.com/srikarchundury/diaq_for_hamsim}.
It implements diagonal-budgeted Trotterization on top of \texttt{libdiaq} and benchmarks it against
Qiskit-Aer and SciPy baselines across all Hamiltonians in Tab.~\ref{tab:benchmarks}.

\textbf{Setup:}

\begin{lstlisting}[frame=single, basicstyle=\small\ttfamily, breaklines=true]
$ git clone --recurse-submodules \
    https://github.com/srikarchundury/diaq_for_hamsim.git
$ cd diaq_for_hamsim
$ pip install -r requirements.txt
\end{lstlisting}

\textbf{Download HamLib datasets (TFIM example):}

\begin{lstlisting}[frame=single, basicstyle=\small\ttfamily, breaklines=true]
$ wget -r -np -nH --cut-dirs=4 -R "index.html*" \
    https://portal.nersc.gov/cfs/m888/dcamps/hamlib/condensedmatter/tfim/
\end{lstlisting}

\textbf{Run end-to-end benchmark:}

\begin{lstlisting}[frame=single, basicstyle=\small\ttfamily, breaklines=true]
$ python benchmark_apps/app2_hamlib_latest.py \
    10 tfim.hdf5 \
    "graph-1D-grid-pbc-qubitnodes_Lx-10_h-6" \
    diaq 1.8 5 True
\end{lstlisting}

The seven positional arguments are:
\texttt{PERCENT\_DIAGONALS} (diagonal budget as a percentage of the maximum possible diagonal count,
e.g., \texttt{10} for 10\%),
\texttt{HDF5\_FILE} (path to a HamLib HDF5 file),
\texttt{HDF5\_KEY} (dataset key identifying the Hamiltonian instance, matching the
\textbf{HDF5File/Key} column of Tab.~\ref{tab:benchmarks_large}),
\texttt{METHOD} (one of \texttt{diaq}, \texttt{diaq-gpu}, \texttt{csr},
\texttt{expm\_multiply}, or \texttt{qiskit-aer}),
\texttt{FINAL\_TIME} (total evolution time $t$),
\texttt{ITRS} (timing repetitions for averaging),
and \texttt{TEST\_FIDELITY} (\texttt{True} to compute fidelity against a dense
NumPy reference).

%% file: tables/benchmarks.tex
\begin{table*}[!htbp]
  \centering
  \small
  \caption{HamLib benchmarks used in this study; Labels match the figures.}
  \label{tab:benchmarks_large}
  \begin{tabularx}{\linewidth}{
      l
      l
      >{\hsize=1.0\hsize\raggedright\arraybackslash}X
      >{\centering\arraybackslash}p{0.9cm}
      >{\centering\arraybackslash}p{1.4cm}
      >{\centering\arraybackslash}p{2.0cm}
    }
      \toprule
      \textbf{Application} & \textbf{Label} & \textbf{HDF5File/Key} & \textbf{Qubits} & \textbf{\#Diagonals} & \textbf{Sparsity (\%)} \\
      \midrule
      \multirow{6}{*}{TSP}
        & TSP-1 & \texttt{\detokenize{TSP_Ncity-4/tsp_rand-008_Ncity-4_enc-stdbinary}} & 8  & 1 & 99.6094 \\
        & TSP-2 & \texttt{\detokenize{TSP_Ncity-4/tsp_rand-007_Ncity-4_enc-stdbinary}} & 8  & 1 & 99.6094 \\
        & TSP-3 & \texttt{\detokenize{TSP_Ncity-5/tsp_rand-007_Ncity-5_enc-stdbinary}} & 15 & 1 & 99.9969 \\
        & TSP-4 & \texttt{\detokenize{TSP_Ncity-5/tsp_rand-006_Ncity-5_enc-stdbinary}} & 15 & 1 & 99.9969 \\
        & TSP-5 & \texttt{\detokenize{TSP_Ncity-4/tsp_rand-008_Ncity-4_enc-unary}}    & 16 & 1 & 99.9985 \\
        & TSP-6 & \texttt{\detokenize{TSP_Ncity-4/tsp_rand-006_Ncity-4_enc-unary}}    & 16 & 1 & 99.9985 \\
      \midrule
      \multirow{7}{*}{Heisenberg}
        & HEIS-1 & \texttt{\detokenize{heis/graph-1D-grid-pbc-qubitnodes_Lx-6_h-5}} & 6  & 13 & 93.7500 \\
        & HEIS-2 & \texttt{\detokenize{heis/graph-1D-grid-pbc-qubitnodes_Lx-6_h-3}} & 6  & 13 & 93.7500 \\
        & HEIS-3 & \texttt{\detokenize{heis/graph-3D-grid-pbc-qubitnodes_Lx-2_Ly-2_Lz-2_h-5}} & 8  & 25 & 97.3145 \\
        & HEIS-4 & \texttt{\detokenize{heis/graph-3D-grid-pbc-qubitnodes_Lx-2_Ly-2_Lz-2_h-3}} & 8  & 25 & 97.3267 \\
        & HEIS-5 & \texttt{\detokenize{heis/graph-1D-grid-nonpbc-qubitnodes_Lx-10_h-5}} & 10 & 19 & 99.4629 \\
        & HEIS-6 & \texttt{\detokenize{heis/graph-1D-grid-nonpbc-qubitnodes_Lx-10_h-3}} & 10 & 19 & 99.4629 \\
        & HEIS-7 & \texttt{\detokenize{heis/graph-1D-grid-nonpbc-qubitnodes_Lx-14_h-0.5}} & 14 & 27 & 99.9548 \\
      \midrule
      \multirow{8}{*}{MaxCut}
        & MAXCUT-1 & \texttt{\detokenize{ham-graph-regular_reg-4/reg-4_n-10_rinst-16}} & 10 & 1 & 99.9390 \\
        & MAXCUT-2 & \texttt{\detokenize{ham-graph-regular_reg-4/reg-4_n-10_rinst-17}} & 10 & 1 & 99.9386 \\
        & MAXCUT-3 & \texttt{\detokenize{ham-graph-regular_reg-4/reg-4_n-12_rinst-18}} & 12 & 1 & 99.9840 \\
        & MAXCUT-4 & \texttt{\detokenize{ham-graph-regular_reg-4/reg-4_n-12_rinst-16}} & 12 & 1 & 99.9834 \\
        & MAXCUT-5 & \texttt{\detokenize{ham-graph-regular_reg-4/reg-4_n-14_rinst-17}} & 14 & 1 & 99.9957 \\
        & MAXCUT-6 & \texttt{\detokenize{ham-graph-regular_reg-4/reg-4_n-14_rinst-18}} & 14 & 1 & 99.9958 \\
        & MAXCUT-7 & \texttt{\detokenize{ham-graph-regular_reg-4/reg-4_n-16_rinst-16}} & 16 & 1 & 99.9989 \\
        & MAXCUT-8 & \texttt{\detokenize{ham-graph-regular_reg-4/reg-4_n-16_rinst-19}} & 16 & 1 & 99.9989 \\
      \midrule
      \multirow{6}{*}{TFIM}
        & TFIM-1 & \texttt{\detokenize{tfim/graph-1D-grid-pbc-qubitnodes_Lx-7_h-6}}  & 7  & 15 & 93.7500 \\
        & TFIM-2 & \texttt{\detokenize{tfim/graph-3D-grid-pbc-qubitnodes_Lx-2_Ly-2_Lz-2_h-6}} & 8  & 17 & 96.5820 \\
        & TFIM-3 & \texttt{\detokenize{tfim/graph-2D-triag-pbc-qubitnodes_Lx-3_Ly-5_h-6}}     & 9  & 19 & 98.0469 \\
        & TFIM-4 & \texttt{\detokenize{tfim/graph-1D-grid-pbc-qubitnodes_Lx-10_h-6}}           & 10 & 21 & 98.9258 \\
        & TFIM-5 & \texttt{\detokenize{tfim/graph-1D-grid-nonpbc-qubitnodes_Lx-14_h-0.5}}           & 14 & 29 & 99.9084 \\
        & TFIM-6 & \texttt{\detokenize{tfim/graph-1D-grid-nonpbc-qubitnodes_Lx-16_h-0.1}}         & 16 & 33  & 99.9741 \\
      \bottomrule
  \end{tabularx}
\end{table*}